\documentclass[10pt,a4paper]{article}
\usepackage{fullpage}

\usepackage[scaled]{helvet}
\usepackage{graphicx}
\usepackage{xcolor}
\usepackage{multirow}
\usepackage[utf8]{inputenc}
\usepackage[round]{natbib}

\usepackage{upgreek}

\usepackage{tikz}
\usepackage[firstpage=true]{background}
\SetBgScale{1}
\SetBgAngle{0}
\SetBgOpacity{1}
\SetBgContents{\begin{tikzpicture}[remember picture,overlay]
        \node[anchor=north east,shift={(-1,-0.6)}]
            at (current page.north east) {
            \orighref{https://doi.org/10.24072/pci.mcb.100183}{\includegraphics[height=3cm]{{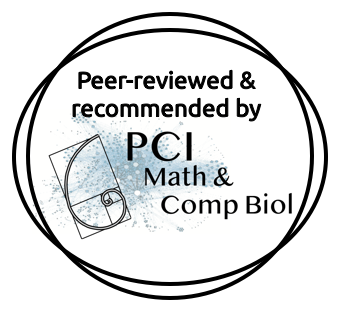}}}};
\end{tikzpicture}}

\usepackage{url,hyperref,microtype}
\usepackage{xspace}
\usepackage[onehalfspacing]{setspace}

\hypersetup{
  breaklinks=true,  
  colorlinks=true,
  urlcolor=red,
  linkcolor=blue,
  citecolor=blue,
}
\usepackage{fontawesome}
\let\orighref\href
\renewcommand{\href}[2]{\orighref{#1}{#2\,\faExternalLink}}
\renewcommand{\url}[1]{\href{#1}{#1}}

\newcommand{\doi}[1]{doi:\href{http://dx.doi.org/#1}{\nolinkurl{#1}}}

    \makeatletter
    \def\maxwidth{\ifdim\Gin@nat@width>\linewidth\linewidth
    \else\Gin@nat@width\fi}
    \makeatother

    \usepackage{adjustbox} 
    \usepackage{xcolor} 
    \usepackage{enumerate} 
    \usepackage{amsmath} 
    \usepackage{amssymb} 
    \usepackage{textcomp} 
    \AtBeginDocument{%
        \def\PYZsq{\textquotesingle}
    }
    \usepackage{upquote} 
    \usepackage{eurosym} 
    \usepackage[mathletters]{ucs} 
    \usepackage{fancyvrb} 
    \usepackage{grffile} 
    \usepackage{hyperref}
    \usepackage{longtable} 
    \usepackage{booktabs}  

    \usepackage[inline]{enumitem} 
    \usepackage[normalem]{ulem} 
\usepackage{url,microtype,subcaption}

    \definecolor{ansi-black}{HTML}{3E424D}
    \definecolor{ansi-black-intense}{HTML}{282C36}
    \definecolor{ansi-red}{HTML}{E75C58}
    \definecolor{ansi-red-intense}{HTML}{B22B31}
    \definecolor{ansi-green}{HTML}{00A250}
    \definecolor{ansi-green-intense}{HTML}{007427}
    \definecolor{ansi-yellow}{HTML}{DDB62B}
    \definecolor{ansi-yellow-intense}{HTML}{B27D12}
    \definecolor{ansi-blue}{HTML}{208FFB}
    \definecolor{ansi-blue-intense}{HTML}{0065CA}
    \definecolor{ansi-magenta}{HTML}{D160C4}
    \definecolor{ansi-magenta-intense}{HTML}{A03196}
    \definecolor{ansi-cyan}{HTML}{60C6C8}
    \definecolor{ansi-cyan-intense}{HTML}{258F8F}
    \definecolor{ansi-white}{HTML}{C5C1B4}
    \definecolor{ansi-white-intense}{HTML}{A1A6B2}

    \providecommand{\tightlist}{%
      \setlength{\itemsep}{0pt}\setlength{\parskip}{0pt}}
    \DefineVerbatimEnvironment{Highlighting}{Verbatim}{commandchars=\\\{\}}
    \newenvironment{Shaded}{}{}
    
    \newcommand{\DataTypeTok}[1]{\textcolor[rgb]{0.56,0.13,0.00}{{#1}}}

    \newcommand{\StringTok}[1]{\textcolor[rgb]{0.25,0.44,0.63}{{#1}}}

    \newcommand{\NormalTok}[1]{{#1}}
    
    \newcommand{\ExtensionTok}[1]{{#1}}
    
    \newcommand{\AttributeTok}[1]{\textcolor[rgb]{0.49,0.56,0.16}{{#1}}}

\makeatletter
\def\PY@reset{\let\PY@it=\relax \let\PY@bf=\relax%
    \let\PY@ul=\relax \let\PY@tc=\relax%
    \let\PY@bc=\relax \let\PY@ff=\relax}
\def\PY@tok#1{\csname PY@tok@#1\endcsname}
\def\PY@toks#1+{\ifx\relax#1\empty\else%
    \PY@tok{#1}\expandafter\PY@toks\fi}
\def\PY@do#1{\PY@bc{\PY@tc{\PY@ul{%
    \PY@it{\PY@bf{\PY@ff{#1}}}}}}}
\def\PY#1#2{\PY@reset\PY@toks#1+\relax+\PY@do{#2}}

\@namedef{PY@tok@w}{\def\PY@tc##1{\textcolor[rgb]{0.73,0.73,0.73}{##1}}}
\@namedef{PY@tok@c}{\let\PY@it=\textit\def\PY@tc##1{\textcolor[rgb]{0.24,0.48,0.48}{##1}}}
\@namedef{PY@tok@cp}{\def\PY@tc##1{\textcolor[rgb]{0.61,0.40,0.00}{##1}}}
\@namedef{PY@tok@k}{\let\PY@bf=\textbf\def\PY@tc##1{\textcolor[rgb]{0.00,0.50,0.00}{##1}}}
\@namedef{PY@tok@kp}{\def\PY@tc##1{\textcolor[rgb]{0.00,0.50,0.00}{##1}}}
\@namedef{PY@tok@kt}{\def\PY@tc##1{\textcolor[rgb]{0.69,0.00,0.25}{##1}}}
\@namedef{PY@tok@o}{\def\PY@tc##1{\textcolor[rgb]{0.40,0.40,0.40}{##1}}}
\@namedef{PY@tok@ow}{\let\PY@bf=\textbf\def\PY@tc##1{\textcolor[rgb]{0.67,0.13,1.00}{##1}}}
\@namedef{PY@tok@nb}{\def\PY@tc##1{\textcolor[rgb]{0.00,0.50,0.00}{##1}}}
\@namedef{PY@tok@nf}{\def\PY@tc##1{\textcolor[rgb]{0.00,0.00,1.00}{##1}}}
\@namedef{PY@tok@nc}{\let\PY@bf=\textbf\def\PY@tc##1{\textcolor[rgb]{0.00,0.00,1.00}{##1}}}
\@namedef{PY@tok@nn}{\let\PY@bf=\textbf\def\PY@tc##1{\textcolor[rgb]{0.00,0.00,1.00}{##1}}}
\@namedef{PY@tok@ne}{\let\PY@bf=\textbf\def\PY@tc##1{\textcolor[rgb]{0.80,0.25,0.22}{##1}}}
\@namedef{PY@tok@nv}{\def\PY@tc##1{\textcolor[rgb]{0.10,0.09,0.49}{##1}}}
\@namedef{PY@tok@no}{\def\PY@tc##1{\textcolor[rgb]{0.53,0.00,0.00}{##1}}}
\@namedef{PY@tok@nl}{\def\PY@tc##1{\textcolor[rgb]{0.46,0.46,0.00}{##1}}}
\@namedef{PY@tok@ni}{\let\PY@bf=\textbf\def\PY@tc##1{\textcolor[rgb]{0.44,0.44,0.44}{##1}}}
\@namedef{PY@tok@na}{\def\PY@tc##1{\textcolor[rgb]{0.41,0.47,0.13}{##1}}}
\@namedef{PY@tok@nt}{\let\PY@bf=\textbf\def\PY@tc##1{\textcolor[rgb]{0.00,0.50,0.00}{##1}}}
\@namedef{PY@tok@nd}{\def\PY@tc##1{\textcolor[rgb]{0.67,0.13,1.00}{##1}}}
\@namedef{PY@tok@s}{\def\PY@tc##1{\textcolor[rgb]{0.73,0.13,0.13}{##1}}}
\@namedef{PY@tok@sd}{\let\PY@it=\textit\def\PY@tc##1{\textcolor[rgb]{0.73,0.13,0.13}{##1}}}
\@namedef{PY@tok@si}{\let\PY@bf=\textbf\def\PY@tc##1{\textcolor[rgb]{0.64,0.35,0.47}{##1}}}
\@namedef{PY@tok@se}{\let\PY@bf=\textbf\def\PY@tc##1{\textcolor[rgb]{0.67,0.36,0.12}{##1}}}
\@namedef{PY@tok@sr}{\def\PY@tc##1{\textcolor[rgb]{0.64,0.35,0.47}{##1}}}
\@namedef{PY@tok@ss}{\def\PY@tc##1{\textcolor[rgb]{0.10,0.09,0.49}{##1}}}
\@namedef{PY@tok@sx}{\def\PY@tc##1{\textcolor[rgb]{0.00,0.50,0.00}{##1}}}
\@namedef{PY@tok@m}{\def\PY@tc##1{\textcolor[rgb]{0.40,0.40,0.40}{##1}}}
\@namedef{PY@tok@gh}{\let\PY@bf=\textbf\def\PY@tc##1{\textcolor[rgb]{0.00,0.00,0.50}{##1}}}
\@namedef{PY@tok@gu}{\let\PY@bf=\textbf\def\PY@tc##1{\textcolor[rgb]{0.50,0.00,0.50}{##1}}}
\@namedef{PY@tok@gd}{\def\PY@tc##1{\textcolor[rgb]{0.63,0.00,0.00}{##1}}}
\@namedef{PY@tok@gi}{\def\PY@tc##1{\textcolor[rgb]{0.00,0.52,0.00}{##1}}}
\@namedef{PY@tok@gr}{\def\PY@tc##1{\textcolor[rgb]{0.89,0.00,0.00}{##1}}}
\@namedef{PY@tok@ge}{\let\PY@it=\textit}
\@namedef{PY@tok@gs}{\let\PY@bf=\textbf}
\@namedef{PY@tok@gp}{\let\PY@bf=\textbf\def\PY@tc##1{\textcolor[rgb]{0.00,0.00,0.50}{##1}}}
\@namedef{PY@tok@go}{\def\PY@tc##1{\textcolor[rgb]{0.44,0.44,0.44}{##1}}}
\@namedef{PY@tok@gt}{\def\PY@tc##1{\textcolor[rgb]{0.00,0.27,0.87}{##1}}}
\@namedef{PY@tok@err}{\def\PY@bc##1{{\setlength{\fboxsep}{\string -\fboxrule}\fcolorbox[rgb]{1.00,0.00,0.00}{1,1,1}{\strut ##1}}}}
\@namedef{PY@tok@kc}{\let\PY@bf=\textbf\def\PY@tc##1{\textcolor[rgb]{0.00,0.50,0.00}{##1}}}
\@namedef{PY@tok@kd}{\let\PY@bf=\textbf\def\PY@tc##1{\textcolor[rgb]{0.00,0.50,0.00}{##1}}}
\@namedef{PY@tok@kn}{\let\PY@bf=\textbf\def\PY@tc##1{\textcolor[rgb]{0.00,0.50,0.00}{##1}}}
\@namedef{PY@tok@kr}{\let\PY@bf=\textbf\def\PY@tc##1{\textcolor[rgb]{0.00,0.50,0.00}{##1}}}
\@namedef{PY@tok@bp}{\def\PY@tc##1{\textcolor[rgb]{0.00,0.50,0.00}{##1}}}
\@namedef{PY@tok@fm}{\def\PY@tc##1{\textcolor[rgb]{0.00,0.00,1.00}{##1}}}
\@namedef{PY@tok@vc}{\def\PY@tc##1{\textcolor[rgb]{0.10,0.09,0.49}{##1}}}
\@namedef{PY@tok@vg}{\def\PY@tc##1{\textcolor[rgb]{0.10,0.09,0.49}{##1}}}
\@namedef{PY@tok@vi}{\def\PY@tc##1{\textcolor[rgb]{0.10,0.09,0.49}{##1}}}
\@namedef{PY@tok@vm}{\def\PY@tc##1{\textcolor[rgb]{0.10,0.09,0.49}{##1}}}
\@namedef{PY@tok@sa}{\def\PY@tc##1{\textcolor[rgb]{0.73,0.13,0.13}{##1}}}
\@namedef{PY@tok@sb}{\def\PY@tc##1{\textcolor[rgb]{0.73,0.13,0.13}{##1}}}
\@namedef{PY@tok@sc}{\def\PY@tc##1{\textcolor[rgb]{0.73,0.13,0.13}{##1}}}
\@namedef{PY@tok@dl}{\def\PY@tc##1{\textcolor[rgb]{0.73,0.13,0.13}{##1}}}
\@namedef{PY@tok@s2}{\def\PY@tc##1{\textcolor[rgb]{0.73,0.13,0.13}{##1}}}
\@namedef{PY@tok@sh}{\def\PY@tc##1{\textcolor[rgb]{0.73,0.13,0.13}{##1}}}
\@namedef{PY@tok@s1}{\def\PY@tc##1{\textcolor[rgb]{0.73,0.13,0.13}{##1}}}
\@namedef{PY@tok@mb}{\def\PY@tc##1{\textcolor[rgb]{0.40,0.40,0.40}{##1}}}
\@namedef{PY@tok@mf}{\def\PY@tc##1{\textcolor[rgb]{0.40,0.40,0.40}{##1}}}
\@namedef{PY@tok@mh}{\def\PY@tc##1{\textcolor[rgb]{0.40,0.40,0.40}{##1}}}
\@namedef{PY@tok@mi}{\def\PY@tc##1{\textcolor[rgb]{0.40,0.40,0.40}{##1}}}
\@namedef{PY@tok@il}{\def\PY@tc##1{\textcolor[rgb]{0.40,0.40,0.40}{##1}}}
\@namedef{PY@tok@mo}{\def\PY@tc##1{\textcolor[rgb]{0.40,0.40,0.40}{##1}}}
\@namedef{PY@tok@ch}{\let\PY@it=\textit\def\PY@tc##1{\textcolor[rgb]{0.24,0.48,0.48}{##1}}}
\@namedef{PY@tok@cm}{\let\PY@it=\textit\def\PY@tc##1{\textcolor[rgb]{0.24,0.48,0.48}{##1}}}
\@namedef{PY@tok@cpf}{\let\PY@it=\textit\def\PY@tc##1{\textcolor[rgb]{0.24,0.48,0.48}{##1}}}
\@namedef{PY@tok@c1}{\let\PY@it=\textit\def\PY@tc##1{\textcolor[rgb]{0.24,0.48,0.48}{##1}}}
\@namedef{PY@tok@cs}{\let\PY@it=\textit\def\PY@tc##1{\textcolor[rgb]{0.24,0.48,0.48}{##1}}}

\def\PYZbs{\char`\\}
\def\PYZus{\char`\_}
\def\PYZob{\char`\{}
\def\PYZcb{\char`\}}
\def\PYZca{\char`\^}
\def\PYZam{\char`\&}

\def\PYZgt{\char`\>}
\def\PYZsh{\char`\#}
\def\PYZpc{\char`\%}

\def\PYZhy{\char`\-}
\def\PYZsq{\char`\'}
\def\PYZdq{\char`\"}
\def\PYZti{\char`\~}

\makeatother

    \definecolor{incolor}{rgb}{0.0, 0.0, 0.5}
    \definecolor{outcolor}{rgb}{0.545, 0.0, 0.0}

\begin{document}

\def\Title{Marker and source-marker reprogramming of Most Permissive Boolean networks and ensembles with BoNesis}
\def\Authors{Loïc Paulevé}
\def\Address{$^{}$Univ. Bordeaux, CNRS, Bordeaux INP, LaBRI, UMR 5800, F-33400 Talence, France}
\def\corrAuthor{Loïc Paulev}
\def\corrEmail{loic.pauleve@labri.fr}

\title{\Title}

\author{\Authors}
\date{\footnotesize\Address}

\maketitle

\begin{abstract}
Boolean networks (BNs) are discrete dynamical systems with applications
to the modeling of cellular behaviors. In this paper, we demonstrate how
the software BoNesis can be employed to exhaustively identify
combinations of perturbations which enforce properties on their fixed
points and attractors. We consider marker properties, which specify that
some components are fixed to a specific value. We study 4 variants of
the marker reprogramming problem: the reprogramming of fixed points, of
minimal trap spaces, and of fixed points and minimal trap spaces
reachable from a given initial configuration with the most permissive
update mode. The perturbations consist of fixing a set of components to
a fixed value. They can destroy and create new attractors. In each case,
we give an upper bound on their theoretical computational complexity,
and give an implementation of the resolution using the BoNesis Python
framework. Finally, we lift the reprogramming problems to ensembles of
BNs, as supported by \emph{BoNesis}, bringing insight on possible and
universal reprogramming strategies. This paper can be executed and
modified interactively.
\end{abstract}
\hypertarget{introduction}{%
\section{Introduction}\label{introduction}}

Boolean networks (BNs) are formal discrete dynamical systems with
pertinent applications for modeling cellular differentiation and fate
decision processes
\citep{Saez-Rodriguez2009,Cohen2015,Schwab2021,Zanudo21,Montagud22}. In
these applications, BNs aim at capturing the stable behaviors
(attractors) and the transient dynamics (trajectories) of the cell. From
this perspective, BNs offer a formal framework for predicting
perturbations that destabilize the system and drive it towards a desired
new stable behavior. BN \emph{control} or BN \emph{reprogramming}, in
reference to cellular reprogramming which aims at converting cell types,
is thus receiving a lot of interest from the computational systems
biology community
\citep{Zanudo2015,Yang2018,Biane2018,Mandon2019,Cifuentes2020,Su2020,Rozum2021}.

The reprogramming of BNs led to a range of methods and tools addressing
different instantiations of this problem: with different type of
perturbations (instantaneous, temporary, permanent), different temporal
modalities (one-step, sequential), different scopes (global
reprogramming or from a given initial condition), different restrictions
on the target attractor (fixed points only, attractors of the original
``wild-type'' BN). On top of that, the \emph{update mode} of the BN,
which determines how the trajectories are computed, can play an
important role on the predictions.
In this paper, we address the BN reprogramming with the Most Permissive
(MP) update mode, where attractors are the minimal trap spaces of the BN
\citep{Pauleve2020}. The problems we tackle are related to \emph{marker}
reprogramming: the desired target attractors are specified by a set of
markers, associating a subset of nodes of the network to fixed values
(e.g., \(A=1,C=0\)). After reprogramming, all the configurations in all
(reachable) attractors must be compatible with these markers.
Importantly, the target attractors are not necessarily attractors of the
original (wild-type) BN: the reprogramming can destroy and create new
attractors. In particular, if there is no attractor in the original
model matching with the marker, the reprogramming will identify
perturbations that will create such an attractor and ensure its
reachability. This is a substantial difference with many of the methods
in the literature. Moreover, the approach we present here can return
\emph{all} the possible solutions, possibly up to a given maximum number
of perturbations to apply, and possibly avoiding \emph{uncontrollable}
nodes.

We address the following BN reprogramming problems in the scope of the
MP update mode:

\begin{itemize}
\tightlist
\item
  \emph{P1}: Marker reprogramming of fixed points: after reprogramming,
  all the fixed points of the BN match with the given markers;
  optionally, we can also ensure that at least one fixed point exists.
\item
  \emph{P2}: Source-marker reprogramming of fixed points: after
  reprogramming, all the fixed points that are \emph{reachable from the
  given initial configuration} match with the given markers.
\item
  \emph{P3}: Marker reprogramming of attractors: after reprogramming,
  all the configurations of all the MP attractors (the minimal trap
  spaces) of the BN match with the given markers.
\item
  \emph{P4}: Source-marker reprogramming of attractors: after
  reprogramming, all the configurations of all the attractors that are
  \emph{reachable from the given initial configuration} match with the
  given markers.
\end{itemize}

MP fixed points match with the fixed points of the global Boolean map of
the BN and are thus identical to the fixed points of the (a)synchronous
update modes. MP attractors match with so-called \emph{minimal trap
spaces} of the BN, which are the smallest sub-hypercubes closed by the
Boolean map. Problem \emph{P1} has been already addressed in the
literature, notably by \citet{Biane2018} with the \emph{ActoNet} method
and by \citet{Moon22}, based on bilevel integer programming. To our
knowledge, none of the other problems have been addressed completely in
the literature.

The software \emph{BoNesis}
(\href{https://github.com/bnediction/bonesis}{github.com/bnediction/bonesis})
provides a generic environment for the automated construction of BNs
with MP update mode from specified structural and dynamical properties.
The properties are translated into a logic satisfiability problem,
expressed in Answer-Set Programming (ASP). Initially, \emph{BoNesis} has
been designed for performing BN synthesis \citep{bn-synthesis-ICTAI19}:
solutions of the logic model correspond to BNs that possess the
specified structural and dynamical properties. Leveraging this generic
declarative specification of properties, \emph{BoNesis} is a versatile
tool for reasoning on BNs in general, with the MP update mode: besides
synthesis, it can be used to do model checking, identify fixed points
and attractors in ensemble of BNs, and identifying reprogramming
strategies.

In this paper, we show how the software \emph{BoNesis} can be employed
to solve P1, P2, P3, and P4 in the scope of \emph{locally-monotone} BNs,
where each local function is
\emph{\href{https://en.wikipedia.org/wiki/Unate_function}{unate}}, i.e.,
where each local function never depends on both positively and
negatively from a same component. Locally-monotone BNs cover all the
models where it assumed that a node cannot be both an activator and
inhibitor of a same other node, which is a common assumption when
modeling biological system.

\emph{BoNesis} enables reasoning on \emph{ensembles} of BNs: one of its
basic input is the domain of BNs to consider. This domain could be
reduced to a singleton BN: in that case, the reasoning is similar to
standard model checking and reprogramming. In general, the domain is
specified from an influence graph, possibly with additional constraints
on the underlying logical functions. For BN synthesis, this domain is
used to delimit symbolically the set of candidate models: \emph{BoNesis}
will output the subset of them that verify the desired dynamical
properties. We show how problems P1 to P4 can be partly lifted to
ensembles of BNs using this approach.
The paper is structured as follows. The \emph{Methods} section gives the
necessary background on BNs and MP update mode and formulation of
elementary dynamical properties as satisfaction problems, as well as
main principles of the \emph{BoNesis} environment. The \emph{Results}
section details how the different reprogramming problems P1-P4 can be
addressed using \emph{BoNesis} and shows some experiments to assess
their scalability. Finally, the \emph{Discussion} section sketches
possible extensions of the addressed problems and underlines current
challenges for their resolution.

This paper is \emph{executable}: it contains snippets of Python code
employing \emph{BoNesis} to demonstrate its usage on small examples,
including command line usage. Instructions for its execution are given
at the beginning of the \emph{Results} section. It can be visualized
online at
\href{https://nbviewer.org/github/bnediction/reprogramming-with-bonesis/blob/release/paper.ipynb}{nbviewer.org/github/bnediction/reprogramming-with-bonesis/blob/release/paper.ipynb}
and interactively executed either online at
\href{https://mybinder.org/v2/gh/bnediction/reprogramming-with-bonesis/release}{mybinder.org/v2/gh/bnediction/reprogramming-with-bonesis/release}
using \href{https://mybinder.org/}{mybinder} online free service, or
locally, following instructions given later in this paper.
\hypertarget{methods}{%
\section{Methods}\label{methods}}

\hypertarget{boolean-networks-and-the-most-permissive-update-mode}{%
\subsection{Boolean networks and the Most Permissive update
mode}\label{boolean-networks-and-the-most-permissive-update-mode}}

\hypertarget{basic-definitions}{%
\subsubsection{Basic definitions}\label{basic-definitions}}

A Boolean network (BN) of dimension \(n\) is specified by a function
\(f: \mathbb B^n\to\mathbb B^n\) where \(\mathbb B = \{0,1\}\) is the
Boolean domain. For \(i\in \{1,\cdots,n\}\),
\(f_i:\mathbb B^n\to\mathbb B\) is referred to as the \emph{local
function} of the \emph{component} \(i\). The Boolean vectors
\(x\in\mathbb B^n\) are called \emph{configurations}, where for any
\(i\in\{1,\cdots,n\}\), \(x_i\) denotes the \emph{state} of component
\(i\) in the configuration \(x\).

A BN \(f\) is \emph{locally monotone} whenever every of its local
functions are \emph{unate}: for each \(i\in\{1,\cdots,n\}\), there
exists an ordering of components \(\preceq^i\in \{\leq, \geq\}^n\) such
that \(\forall x,y\in \mathbb B^n\),
\((x_1\preceq^i_1 y_1 \wedge \cdots x_n\preceq^i_n y_n) \implies f_i(x) \leq f_i(y)\).
Intuitively, a BN is locally monotone whenever each of its local
function can be expressed in propositional logic such that each variable
appears either never or always with the same sign. For instance
\(f_1(x) = x_1\vee (\neg x_3 \wedge x_2)\) is unate, whereas
\(f_1(x) = x_2 \oplus x_3 = (x_2\wedge\neg x_3)\vee (\neg x_2\wedge x_3)\)
is not unate.

\paragraph{Example}

The BN \(f\) of dimension \(3\) with \(f_1(x)=\neg x_2\),
\(f_2(x)=\neg x_1\), and \(f_3(x) = \neg x_1\wedge x_2\) is locally
monotone; and an instance of application is \(f(000)=110\).

Locally monotone BNs should not be confused with \emph{monotone} BNs
where a component appears in \emph{all} local functions with the same
sign. Monotone BNs are a particular case of locally-monotone BNs.

\hypertarget{update-modes}{%
\subsubsection{Update modes}\label{update-modes}}

Given a BN \(f\) and a configuration \(x\), the \emph{update mode}
specifies how to compute the next configuration. There is a vast zoo of
update modes \citep{PS22}, but traditionally, two modes are usually
considered in biological modeling: the \emph{synchronous} (or parallel)
deterministic mode, where the next configuration is given by its
application to \(f\) (\(x\) is succeeded by \(f(x)\)), and the
\emph{fully asynchronous} (often denoted only asynchronous) where the
next configuration results from the application of only one local
function, chosen non-deterministically.

However, (a)synchronous update modes do not lead to a complete
qualitative abstraction of quantitative systems and preclude the
prediction of trajectories that are actually feasible when considering
time scales or concentration scales. The \emph{Most Permissive} (MP)
\citep{Pauleve2020,Pauleve2021} is a recently-introduced update mode
which brings the formal guarantee to capture any trajectory that is
feasible by any quantitative system compatible with the Boolean network
(see (Paulevé et al., 2020) for details). The main idea behind the MP
update mode is to systematically consider a potential delay when a
component changes state, and consider any additional transitions that
could occur if the changing component is in an intermediate state. It
can be modeled as additional \emph{dynamic} states ``increase''
(\(\nearrow\)) and ``decrease'' (\(\searrow\)): when a component can be
activated, it will first go through the ``increase'' state where it can
be interpreted as either 0 or 1 by the other components, until
eventually reaching the Boolean 1 state; and symmetrically for
deactivation. A formal definition of MP dynamics is given later in this
section.

\hypertarget{perturbations}{%
\subsubsection{Perturbations}\label{perturbations}}

In this paper we will consider BN \emph{perturbations} that modify the
local functions of some components so they become a constant function.
Perturbations model mutations, where a gene is silenced or
constitutively activated. Mathematically, a perturbation is a map
associating a set of components to a Boolean value, for instance,
\(P = \{ 2 \mapsto 0, 4 \mapsto 1\}\). Given a perturbation \(P\), the
\emph{perturbed BN} \(f/P\) is given by, for each component
\(i\in \{1,\ldots,n\}\): \[
(f/P)_i(x) = \begin{cases}
b & \text{ if }i \mapsto b \in P\\
f_i(x) & \text{otherwise.}
\end{cases}
\]
\hypertarget{quantified-boolean-expressions-and-computational-complexity}{%
\subsection{Quantified Boolean expressions and computational
complexity}\label{quantified-boolean-expressions-and-computational-complexity}}

A Boolean expression is a logic formula composed of Boolean variables
and propositional logic operators (negation \(\neg\), conjunction
\(\wedge\), disjunction \(\vee\), implication \(\implies\), equivalence
\(\equiv\), exclusive disjunction \(\oplus\)). Given variables
\(x_1,\cdots,x_m\), a \emph{quantified} Boolean expression is of the
form \(Q_1 x_1 \cdots Q_n x_m \phi(x_1, \cdots, x_m)\) where
\(Q_1, \cdots, Q_m\) can be the existential \(\exists\) or universal
\(\forall\) quantifier, and \(\phi(x_1,\cdots,x_m)\) a quantifier-free
Boolean expression composed of variables \(x_1, \cdots, x_m\). For
instance, consider the quantified Boolean expression
``\(\exists x_1\exists x_2 \forall x_3\, (x_3 \wedge x_1) \vee (\neg x_3 \wedge \neg x_2)\)''.
This expression is satisfiable: fix \(x_1=1\) and \(x_2=0\), then the
Boolean expression becomes equivalent to \(x_3 \vee \neg x_3\) which is
true for all assignments of \(x_3\).

Deciding whether such an expression is true (satisfiable) is a
fundamental problem in computer science. The complexity of this problem
actually depends on the alternation of quantifiers. Thus, in the
following we will classify the quantified Boolean expressions by their
sequence of quantifiers \(Q_1\cdots Q_m\) but ignoring repetitions: an
\(\exists\exists\forall\forall\forall\exists\exists\forall\)-expression
has the same decision complexity as an
\(\exists\forall\exists\forall\)-expression.

Computational complexity \citep{Papadimitriou} is a fundamental theory
of computer science to classify decision problems: a (decision) problem
is in class C whenever there exists an algorithm of worst-case
complexity C, C referring to either a time or space complexity. For
instance, the class PTIME gathers all the problems that can be decided
in time polynomial with the size of the input (e.g., the length of the
Boolean expression).

The decision of satisfiability of \(\exists\)-expressions is the
infamous (Boolean) SAT(isfiability) problem, which is NP-complete: it
can be solved by a non-deterministic polynomial time algorithm, and it
is among the hardest problems in this class: any problem in NP can be
(efficiently) transformed into a SAT problem. In practice, our computers
being deterministic, the resolution of the SAT problems employs
algorithms running in worst-case time and space exponential with the
number of variables in the Boolean expression. However, modern SAT
solvers can approach expressions with thousands to millions of
variables.

The decision of satisfiability of \(\forall\)-expressions can be seen as
a complementary problem to \(\exists\)-expression: \(\forall X~\phi(X)\)
is satisfiable if and only if \(\exists X~\neg\phi(X)\) is not
satisfiable: it is a \emph{co}NP-complete problem. It is not known
whether coNP = NP.

Then, the alternation of quantifiers makes the problem climbing into the
so-called polynomial hierarchy\footnote{See
  \href{https://en.wikipedia.org/wiki/Polynomial_hierarchy}{en.wikipedia.org/wiki/Polynomial\_hierarchy}}.
\(\exists\ldots\)-expressions are \(\Sigma_k^{\mathrm P}\)-complete
problems, where \(k\) is the number of alternating quantifiers (starting
with \(\exists\)), while \(\forall\ldots\)-expressions are
\(\Pi_k^{\mathrm P}\)-complete (\(\Sigma_1^{\mathrm P}\)=NP and
\(\Pi_1^{\mathrm P}\)=coNP). It is not known yet whether all these
complexity classes are equal, but in practice, algorithms of resolution
scale rapidly down with their height in the polynomial hierarchy. Each
of these complexity classes are included in PSPACE, the class of
problems solvable in polynomial space. PSPACE-complete problems, such as
the verification of properties of asynchronous BNs, are known to be
difficult to tackle in practice (currently limited to a couple of
hundreds of variables in the case of BNs).

The reader should keep in mind that the length of the expression is a
crucial parameter for the decision complexity. When variables have a
finite domain, one can rewrite quantified Boolean expression in a
universal-free one. However, the length of the obtained expression will
be exponentially larger.

In the rest of the paper, for the sake of simplicity, we will not fully
detail the size of the quantified Boolean expression we derive, and are
expected to be of length linear or polynomial with the size of the BN.
\hypertarget{elementary-dynamical-properties-and-their-complexity}{%
\subsection{Elementary dynamical properties and their
complexity}\label{elementary-dynamical-properties-and-their-complexity}}

We present the formal aspects of the MP dynamics that are employed in
the rest of the paper, i.e., related to attractors and the reachability
of attractors. The proofs and full MP definition and properties can be
found in \citep{Pauleve2020}.

\hypertarget{sub-hypercubes-and-trap-spaces}{%
\subsubsection{Sub-hypercubes and trap
spaces}\label{sub-hypercubes-and-trap-spaces}}

A sub-hypercube specifies for each dimension of the BN if it is either
fixed to a Boolean value, or free: it can be characterized by a vector
\(h\in \{0,1,*\}^n\). Its \emph{vertices} are denoted by
\(c(h) = \{ x\in \mathbb B^n\mid h_i\neq *\implies x_i=h_i\}\). For
instance, \(h=0**\) is a sub-hypercube of dimension 3, with
\(c(h) = \{000, 001, 010, 011\}\).

A sub-hypercube \(h\) is a \emph{trap space} whenever for each of its
vertices \(x\in c(h)\), \(f(x)\) is also one of its vertices (\(h\) is
closed by \(f\)). In particular, the (sub-)hypercube \(\mathbf *_n\) is
always a trap space.

A sub-hypercube \(h\) is \emph{smaller} than a sub-hypercube \(h'\),
denoted by \(h \preceq h'\) whenever \(c(h)\subseteq c(h')\).
Equivalently, this means that each non-free component of \(h'\) is fixed
to the same value in \(h\):
\(h \preceq h' \iff \forall i\in \{1,\ldots,n\}, h'_i\neq *\implies h_i=h'_i\).

\hypertarget{mp-attractors-are-minimal-trap-spaces}{%
\subsubsection{MP attractors are minimal trap
spaces}\label{mp-attractors-are-minimal-trap-spaces}}

The attractors of MP dynamics are the \emph{minimal trap spaces} of the
Boolean function \(f\) \citep{Pauleve2020}, i.e., the trap spaces which
do not include strictly smaller trap spaces. Thus, we denote MP
attractors by sub-hypercubes, i.e., an MP attractor \(A\) is a vector in
\(\{0,1,*\}^n\). Therefore, a component with a \(*\) value in an MP
attractor \(A\) indicates that the component that can always oscillate
between 0 and 1 in the (cyclic) attractor.

The computational complexity of decision problems related to minimal
trap spaces has been extensively addressed in
\citep{TrapSpaceComplexity} with different representations of Boolean
networks. For the case of local functions represented with propositional
logic, as we consider here, deciding whether a sub-hypercube is a trap
space is coNP-complete problem, whereas deciding whether it is a
\emph{minimal} trap space is a coNP\(^\text{coNP}\)-complete problem,
i.e., equivalent to the decision of satisfiability of \(\forall\exists\)
expressions. In the case of \emph{locally-monotone} BNs, deciding
whether a sub-hypercube is a trap space is in PTIME, whereas deciding
whether it is a minimal trap spaces is a coNP-complete problem, i.e.,
equivalent to the decision of satisfiability of \(\forall\)-expressions.

\hypertarget{mp-reachability-of-attractors}{%
\subsubsection{MP reachability of
attractors}\label{mp-reachability-of-attractors}}

Given a configuration \(x\) and an MP attractor \(A\in \{0,1,*\}^n\),
there is an MP trajectory from \(x\) to any configuration \(y\in A\) if
and only if \(A\) is smaller than the smallest trap space
\emph{containing} \(x\). We write \(\operatorname{reach}(x,y)\) the
existence of such a trajectory.

Let us denote by \(\operatorname{TS}(x) \in \{0,1,*\}^n\) the smallest
trap space containing \(x\). The computation of
\(h=\operatorname{TS}(x)\) can be performed from \(x\) by iteratively
freeing the components necessarily to fulfill the closeness property.
Here is a sketch of algorithm, where
\texttt{SAT(h,\ f{[}i{]}\ =\ -x{[}i{]})} is true if and only if there
exists a configuration \(y\in c(h)\) such that \(f_i(y)=\neg x_i\):

\begin{verbatim}
Algorithm TS(x: configuration)
Returns sub-hypercube h
--
h := x
repeat
   changed := false
   for i in 1..n:
      if h[i] != * and SAT(h, f[i] = -x[i]):
          h[i] := *
          changed := true
while changed
\end{verbatim}

In the worst case, this algorithm makes a quadratic number of calls to
the \texttt{SAT} problem. Therefore, the decision of MP reachability of
attractors is in PTIME\(^\text{NP}\) in general\footnote{this problem is
  actually in NP when allowing a number of variables quadratic with
  \(n\)}, and PTIME in the locally-monotone case.

Note that the general MP reachability property is not addressed here,
but its overall complexity is identical. With (a)synchronous update
modes, it is a PSPACE-complete problem.

\hypertarget{belonging-to-an-mp-attractor}{%
\subsubsection{Belonging to an MP
attractor}\label{belonging-to-an-mp-attractor}}

In the following, we will consider the problem of deciding whether a
given configuration \(x\) belongs to an MP attractor of \(f\). We write
\(\operatorname{IN-ATTRACTOR}(x)\) such a property. This can be verified
in two steps: (1) compute the smallest trap spaces containing \(x\),
noted \(\operatorname{TS}(x)\), and (2) verify whether
\(\operatorname{TS}(x)\) is a minimal trap space. This later property is
true if and only if for any vertex \(y\) of \(\operatorname{TS}(x)\),
the minimal trap space containing \(y\) is equal to
\(\operatorname{TS}(x)\): \[
\operatorname{IN-ATTRACTOR}(x) \equiv \forall y \in c(\operatorname{TS}(x)), \operatorname{TS}(y) = \operatorname{TS}(x) \enspace.
\]

Finally, given a set of perturbations \(P\), we write
\(\operatorname{TS}_P(x)\) for the small trap space of perturbed BN
\((f/P)\) containing \(x\), and \(\operatorname{IN-ATTRACTOR}_P(x)\) the
property of \(x\) belonging to an attractor of the perturbed BN
\((f/P)\).
\hypertarget{bonesis}{%
\subsection{BoNesis}\label{bonesis}}

\emph{BoNesis}
(\href{https://github.com/bnediction/bonesis}{github.com/bnediction/bonesis})
is a Python library which has been primarily designed for identifying
BNs satisfying user-given dynamical properties among a given domain of
BNs and with the MP update mode. It takes as input (1) a domain of BNs
\(\mathbb F\), and (2) a set of Boolean dynamical properties \(\phi\),
and can enumerate the BNs \(f \in \mathbb F\) such that
\(f\models \phi\), i.e., \(f\) verifies the properties \(\phi\).

Currently, the domain of BNs \(\mathbb F\) can be one of the following:

\begin{itemize}
\tightlist
\item
  A singleton locally-monotone BN \(\mathbb F=\{f\}\). In that case,
  \emph{BoNesis} can be employed as a model checker to verify that \(f\)
  has the specified dynamical properties. In this paper, this is the
  main setting we will consider, in order to predict perturbations to
  reprogram the attractors of \(f\).
\item
  An explicit ensemble of locally-monotone BNs
  \(\mathbb F=\{ f^1,\cdots, f^m \}\).
\item
  Any locally-monotone BN matching with a given \emph{influence graph}
  \(\mathcal G\): \(\mathbb F = \{ f\mid G(f)\subseteq \mathcal G\}\).
  An influence graph is a signed digraph between components, i.e., of
  the form \((\{1,\cdots,n\},V)\) with
  \(V\subseteq \{1,\cdots,n\}\times \{+1,-1\}\times \{1,\cdots n\}\).
  The influence graph of a BN \(f\), denoted by \(G(f)\) has an edge
  \(i\xrightarrow{s} j\) if and only there exists a configuration
  \(x\in\mathbb B^n\) such that
  \(f_j(x_1, \ldots, x_{i-1}, 1, x_{i+1},\ldots, x_n) - f_j(x_1, \ldots, x_{i-1}, 0, x_{i+1},\ldots, x_n) = s\).
\item
  Any locally-monotone BN matching with a partially-defined BN following
  the AEON framework \citep{Benes2021}.
\end{itemize}

\emph{BoNesis} offers a Python programming interface to declare the
dynamical properties over BNs, including reachability, fixed points and
trap spaces. \emph{BoNesis} relies on Answer-Set Programming (ASP) and
the ASP solver \href{https://potassco.org/clingo}{\emph{clingo}} for the
enumeration of solutions. ASP is a declarative logic programming
framework for expressing combinatorial decision problems and enumerate
their solutions, possibly with optimizations. ASP can be employed for
efficiently solving \(\exists\)- and \(\exists\forall\)-expressions,
thus having an expressiveness higher than classical SAT.

We emphasize that \emph{BoNesis} is currently restricted to
locally-monotone BNs only for which efficient logical encoding of
domains of models are possible. Whereas it is a common assumption when
modeling of biological systems (a node cannot be both an activator and
inhibitor of a same other node), non-monotone BNs are also employed, and
cannot be addressed with the current implementation.

The usage of \emph{BoNesis} Python programming interface and command
line will be explained along with the code snippets provided in the next
sections.
\hypertarget{results}{%
\section{Results}\label{results}}

We show how the general declarative approach of \emph{BoNesis} can be
instantiated to compute the complete solutions to the P1, P2, P3, and P4
reprogramming problems on BNs, and also extend the reasoning to
ensembles of BNs. Importantly, note that \emph{BoNesis} currently
supports only locally-monotone BNs.

This is an \emph{executable paper} which demonstrates the use of
\emph{BoNesis} for the reprogramming of BNs. The corresponding notebook
can be downloaded from
\href{https://nbviewer.org/github/bnediction/reprogramming-with-bonesis/blob/release/paper.ipynb}{nbviewer.org/github/bnediction/reprogramming-with-bonesis/blob/release/paper.ipynb}.
Its execution requires the \href{https://jupyter.org/}{Jupyter notebook}
system, \href{https://python.org}{Python}, and the Python package
\texttt{bonesis} to be installed, see
\href{https://github.com/bnediction/bonesis}{github.com/bnediction/bonesis}
for instructions. Alternatively, the notebook can be executed within the
CoLoMoTo Docker distribution \citep{ColomotoDocker}, using the Docker
image \texttt{colomoto/colomoto-docker:2023-02-01}, which can be
launched as follows:

\begin{verbatim}
pip install colomoto-docker
colomoto-docker -V 2023-03-01
\end{verbatim}

then open \url{http://127.0.0.1:8888} and upload the notebook from the
Jupyter interface. See
\href{https://github.com/bnediction/reprogramming-with-bonesis}{github.com/bnediction/reprogramming-with-bonesis}
for further help.
\begin{small}
\begin{Verbatim}[commandchars=\\\{\}]
{\color{incolor}In [{\color{incolor}1}]:} \PY{c+ch}{\PYZsh{}!pip install \PYZhy{}\PYZhy{}user bonesis  \PYZsh{} uncomment to install bonesis}
        \PY{k+kn}{import} \PY{n+nn}{bonesis}
\end{Verbatim}
\end{small}

\begin{small}
\begin{Verbatim}[commandchars=\\\{\}]
{\color{incolor}In [{\color{incolor}2}]:} \PY{k+kn}{from} \PY{n+nn}{colomoto\PYZus{}jupyter} \PY{k+kn}{import} \PY{n}{tabulate} \PY{c+c1}{\PYZsh{} for display}
        \PY{k+kn}{import} \PY{n+nn}{pandas} \PY{k}{as} \PY{n+nn}{pd} \PY{c+c1}{\PYZsh{} for display}
        \PY{k+kn}{import} \PY{n+nn}{mpbn} \PY{c+c1}{\PYZsh{} for analyzing individual Boolean networks with MP update mode}
        \PY{k+kn}{from} \PY{n+nn}{colomoto}\PY{n+nn}{.}\PY{n+nn}{minibn} \PY{k+kn}{import} \PY{n}{BooleanNetwork}
\end{Verbatim}
\end{small}

Alternatively, the computation of reprogramming perturbations from
single Boolean networks can be performed using the command line
program\texttt{bonesis-reprogramming}, provided alongside the
\texttt{bonesis} Python package. We detail its usage in each case.
\hypertarget{marker-reprogramming-of-boolean-networks}{%
\subsection{Marker reprogramming of Boolean
networks}\label{marker-reprogramming-of-boolean-networks}}

We first consider the reprogramming of a single BN \(f\) of dimension
\(n\). In the framework of \emph{BoNesis}, this means the domain of BNs
is the singleton \(\mathbb F = \{ f \}\).

A marker \(M\) is a map associating a subset of components of \(f\) to a
Boolean value. For instance, \(M = \{ 1\mapsto 0, 3\mapsto 1\}\) is the
marker where component \(1\) is \(0\) and component \(3\) is \(1\). We
denote by \(dom(M)\) the domain of the map \(M\), i.e., in our example
\(dom(M) = \{ 1, 3\}\). Given a configuration \(x\in \mathbb B^n\), we
say \(x\) matches with a marker \(M\), denoted by \(x\models M\), if and
only if \(\forall i\in dom(M), x_i=M(i)\). Given a set of configurations
\(A\subseteq \mathbb B^n\), we say \(A\) matches with a marker \(M\) if
and only if each of its configurations match with \(M\)
(\(\forall x\in A, x\models M\)). Given \(k\in\mathbb N\), we denote by
\(\mathbb M^{\leq k}\) the sets of maps associating at most \(k\)
components among \(\{1, \cdots, n\}\) to a Boolean value.

The objective of marker-reprogramming is to identify perturbations so
that all the fixed points/attractors of the perturbed \(f\) match with
the marker \(M\). The source-marker reprogramming then focuses on the
fixed points/attractors reachable from a given initial configuration
only, thus potentially requiring fewer perturbations.

A very important aspect of marker reprogramming is that it accounts for
the creation and deletion of attractors due to the perturbation. Thus,
in general, the attractors of the reprogrammed BN are different from the
attractors of the input (wild-type) BN.

In this section, we tackle the following instantiations of the
reprogramming problem:

\begin{enumerate}
\def\labelenumi{\arabic{enumi}.}
\tightlist
\item
  Marker reprogramming of fixed points (\emph{P1});
\item
  Source-marker reprogramming of fixed points (\emph{P2});
\item
  Marker reprogramming of attractors (\emph{P3});
\item
  Source-marker reprogramming of attractors (\emph{P4}).
\end{enumerate}

In each case, we briefly study the complexity of the associated decision
problem (existence of a perturbation given the desired reprogramming
property), and give the Python and command line recipe to identify the
perturbations with \emph{BoNesis}. The following table summarizes the
results, with the complexity in the locally-monotone case and command
line usage:

\begin{longtable}[]{@{}cll@{}}
\toprule()
Problem & Complexity & Command line \\
\midrule()
\endhead
P1 & \(\exists\forall\) & \texttt{{[}base{]}\ -\/-fixpoints} \\
P2 & \(\exists\forall\) &
\texttt{{[}base{]}\ -\/-fixpoints\ -\/-reachable-from\ z} \\
P3 & \(\exists\forall\exists\) & \texttt{{[}base{]}} \\
P4 & \(\exists\forall\exists\) &
\texttt{{[}base{]}\ -\/-reachable-from\ z} \\
\bottomrule()
\end{longtable}

where \texttt{{[}base{]}} is the command line
\texttt{bonesis-reprogramming\ model.bnet\ M\ k}, with
\texttt{model.bnet} the path to a file in
\href{http://colomoto.org/biolqm/doc/format-bnet.html}{BooleanNet
format}, \texttt{M} specifies the marker as a JSON map, \texttt{k} is
the maximum number of simultaneous perturbations, and \texttt{z} is the
initial configuration as a JSON map. For instance,

\begin{Shaded}
\begin{Highlighting}[]
\ExtensionTok{bonesis{-}reprogramming}\NormalTok{ model.bnet }\StringTok{\textquotesingle{}\{"A": 0, "C": 1\}\textquotesingle{}}\NormalTok{ 3 }\DataTypeTok{\textbackslash{}}
        \AttributeTok{{-}{-}reachable{-}from} \StringTok{\textquotesingle{}\{"A":1, "B":0, "C":0,"D":0\}\textquotesingle{}}
\end{Highlighting}
\end{Shaded}
\hypertarget{marker-reprogramming-of-fixed-points-p1}{%
\subsubsection{Marker-reprogramming of fixed points
(P1)}\label{marker-reprogramming-of-fixed-points-p1}}

We identify the perturbations \(P\) of at most \(k\) components so that
all the fixed points of \(f/P\) match with the given marker \(M\). The
associated decision problem can be expressed as the following
\(\exists\forall\)-expression, hence being at most in
\(\Sigma_2^{\mathrm P}\):

\begin{equation}
\exists P\in \mathbb M^{\leq k}, \forall x\in\mathbb B^n, (f/P)(x) = x \Rightarrow x\models M
\end{equation}

``There exists a perturbation being a map of at most \(k\) components to
a Boolean value, such that for all configurations \(x\in\mathbb B^n\),
if \(x\) is a fixed point of the perturbed BN \((f/P)\), then \(x\)
matches with the marker \(M\)''.

Remark that any BN having no fixed point verify the above equation with
an empty perturbation. Thus, in practice, one may also expect that the
perturbed BN possesses at least one fixed point:

\begin{equation}
\exists P\in \mathbb M^{\leq k}, \exists y\in\mathbb B^n, (f/P)(y) = y, \forall x\in\mathbb B^n, (f/P)(x) = x \Rightarrow x\models M\enspace.
\end{equation}

With the \emph{BoNesis} Python interface, this reprogramming property
can be declared as follows, where \texttt{f} is a BN, \texttt{M} the
marker (specified as Python dictionary associating a subset of
components to a Boolean value), and \texttt{k} the maximum number of
components that can be perturbed (at most \(n\)):
\begin{small}
\begin{Verbatim}[commandchars=\\\{\}]
{\color{incolor}In [{\color{incolor}3}]:} \PY{k}{def} \PY{n+nf}{marker\PYZus{}reprogramming\PYZus{}fixpoints}\PY{p}{(}\PY{n}{f}\PY{p}{:} \PY{n}{BooleanNetwork}\PY{p}{,} 
                                           \PY{n}{M}\PY{p}{:} \PY{n+nb}{dict}\PY{p}{[}\PY{n+nb}{str}\PY{p}{,}\PY{n+nb}{bool}\PY{p}{]}\PY{p}{,}
                                           \PY{n}{k}\PY{p}{:} \PY{n+nb}{int}\PY{p}{,} \PY{n}{ensure\PYZus{}exists}\PY{p}{:}\PY{n+nb}{bool}\PY{o}{=}\PY{k+kc}{True}\PY{p}{)}\PY{p}{:}
            \PY{c+c1}{\PYZsh{} f: Boolean network; M: marker; k: maximum number of components to perturb}
            \PY{n}{bo} \PY{o}{=} \PY{n}{bonesis}\PY{o}{.}\PY{n}{BoNesis}\PY{p}{(}\PY{n}{f}\PY{p}{)}
            \PY{n}{P} \PY{o}{=} \PY{n}{bo}\PY{o}{.}\PY{n}{Some}\PY{p}{(}\PY{n}{max\PYZus{}size}\PY{o}{=}\PY{n}{k}\PY{p}{)} \PY{c+c1}{\PYZsh{} perturbations to identify}
            \PY{k}{with} \PY{n}{bo}\PY{o}{.}\PY{n}{mutant}\PY{p}{(}\PY{n}{P}\PY{p}{)}\PY{p}{:}
                \PY{c+c1}{\PYZsh{} impose that all the fixed points of the perturbed BN match with M}
                \PY{n}{bo}\PY{o}{.}\PY{n}{all\PYZus{}fixpoints}\PY{p}{(}\PY{n}{bo}\PY{o}{.}\PY{n}{obs}\PY{p}{(}\PY{n}{M}\PY{p}{)}\PY{p}{)}
                \PY{k}{if} \PY{n}{ensure\PYZus{}exists}\PY{p}{:}
                    \PY{c+c1}{\PYZsh{} impose the existence of at least one fixed point matching with M}
                    \PY{n}{bo}\PY{o}{.}\PY{n}{fixed}\PY{p}{(}\PY{o}{\PYZti{}}\PY{n}{bo}\PY{o}{.}\PY{n}{obs}\PY{p}{(}\PY{n}{M}\PY{p}{)}\PY{p}{)} 
            \PY{k}{return} \PY{n}{P}\PY{o}{.}\PY{n}{assignments}\PY{p}{(}\PY{p}{)}
\end{Verbatim}
\end{small}

The line \texttt{bo\ =\ bonesis.BoNesis(f)} instantiates a
\emph{BoNesis} object \texttt{bo} with the domain \texttt{f}. In this
section, we assume that \texttt{f} is a single Boolean network. It can
be either the path to a BooleanNet file, a \texttt{minibn} object, for
instance as returned by the \texttt{biolqm.to\_minibn} function for
importing models from various formats, or a Python dictionary,
associating components to a Boolean function. Examples are given below.
Then, the line \texttt{P\ =\ bo.Some(max\_size=k)} declares a variable
that will consist of a map associating at most \texttt{k} components to
a Boolean value (the perturbation to be identified). The instruction
\texttt{with\ bo.mutant(P)} opens a context where dynamical properties
will be verified against the BN \texttt{f} with the perturbation
\texttt{P} applied. Within this mutant context, we declare with
\texttt{bo.all\_fixpoints(bo.obs(M))} that each fixed point of the
perturbed model matches with \texttt{M}. Moreover, whenever
\texttt{ensure\_exists} is true, the constraint
\texttt{bo.fixed(\textasciitilde{}bo.obs(M))} imposes that the
configuration associated with \texttt{M}
(\texttt{\textasciitilde{}bo.obs(M)}) is a fixed point. Finally,
\texttt{P.assignments()} returns an iterator over all the possible
assignments of \texttt{P} that fulfill the above dynamical properties.

The corresponding command line is of the form

\begin{Shaded}
\begin{Highlighting}[]
\ExtensionTok{bonesis{-}reprogramming}\NormalTok{ model.bnet M k }\AttributeTok{{-}{-}fixpoints}
\end{Highlighting}
\end{Shaded}

where \texttt{model.bnet} is a BN encoded in BooleanNet format,
\texttt{M} specifies the marker as a JSON map, and \texttt{k} is the
maximum number of perturbations. The existence of at least one fixed
point can be lifted with the option \texttt{-\/-allow-no-fixpoint}.
\paragraph{Example}

We illustrate the marker-reprogramming of fixed points on a small toy BN
example, which has no fixed point. In the following, we use
\texttt{colomoto.minibn.BooleanNetwork} to define this BN.
\begin{small}
\begin{Verbatim}[commandchars=\\\{\}]
{\color{incolor}In [{\color{incolor}4}]:} \PY{n}{f} \PY{o}{=} \PY{n}{BooleanNetwork}\PY{p}{(}\PY{p}{\PYZob{}}
            \PY{l+s+s2}{\PYZdq{}}\PY{l+s+s2}{A}\PY{l+s+s2}{\PYZdq{}}\PY{p}{:} \PY{l+s+s2}{\PYZdq{}}\PY{l+s+s2}{B}\PY{l+s+s2}{\PYZdq{}}\PY{p}{,}
            \PY{l+s+s2}{\PYZdq{}}\PY{l+s+s2}{B}\PY{l+s+s2}{\PYZdq{}}\PY{p}{:} \PY{l+s+s2}{\PYZdq{}}\PY{l+s+s2}{!A}\PY{l+s+s2}{\PYZdq{}}\PY{p}{,}
            \PY{l+s+s2}{\PYZdq{}}\PY{l+s+s2}{C}\PY{l+s+s2}{\PYZdq{}}\PY{p}{:} \PY{l+s+s2}{\PYZdq{}}\PY{l+s+s2}{!A \PYZam{} B}\PY{l+s+s2}{\PYZdq{}}
        \PY{p}{\PYZcb{}}\PY{p}{)}
        \PY{n}{f}
\end{Verbatim}
\end{small}

\begin{small}
\begin{Verbatim}[commandchars=\\\{\}]
{\color{outcolor}Out[{\color{outcolor}4}]:} A <- B
        B <- !A
        C <- !A\&B
\end{Verbatim}
\end{small}
This example BN has two components in negative feedback: they will
oscillate forever. The state of the third component \texttt{C} is then
determined by the state of the oscillating components. The following
command returns its influence graph:
\begin{small}
\begin{Verbatim}[commandchars=\\\{\}]
{\color{incolor}In [{\color{incolor}5}]:} \PY{n}{f}\PY{o}{.}\PY{n}{influence\PYZus{}graph}\PY{p}{(}\PY{p}{)}
\end{Verbatim}
\end{small}

\begin{figure}[tbp]
    \centering
    \includegraphics[scale=0.7]{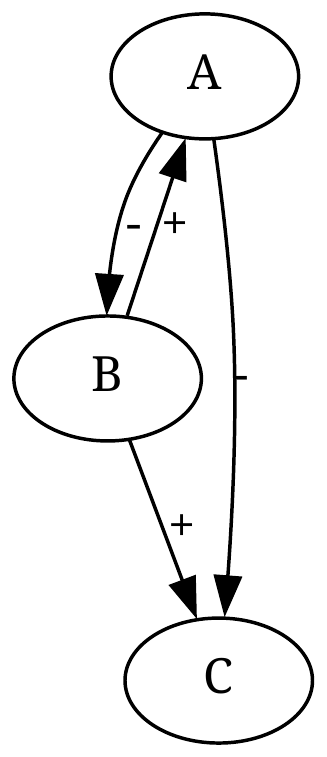}\caption{Influence graph of the example BN for P1}\label{fig:paper_files/paper_19_1.pdf}
    \end{figure}

    \noindent
    The resulting graphics is reproduced in Figure~\ref{fig:paper_files/paper_19_1.pdf}.

With the (fully) asynchronous update mode, the system has a single
attractor, consisting of all the configurations of the network.
\begin{small}
\begin{Verbatim}[commandchars=\\\{\}]
{\color{incolor}In [{\color{incolor}6}]:} \PY{n}{f}\PY{o}{.}\PY{n}{dynamics}\PY{p}{(}\PY{l+s+s2}{\PYZdq{}}\PY{l+s+s2}{asynchronous}\PY{l+s+s2}{\PYZdq{}}\PY{p}{)}
\end{Verbatim}
\end{small}

\begin{figure}[tbp]
    \centering
    \includegraphics[scale=0.7]{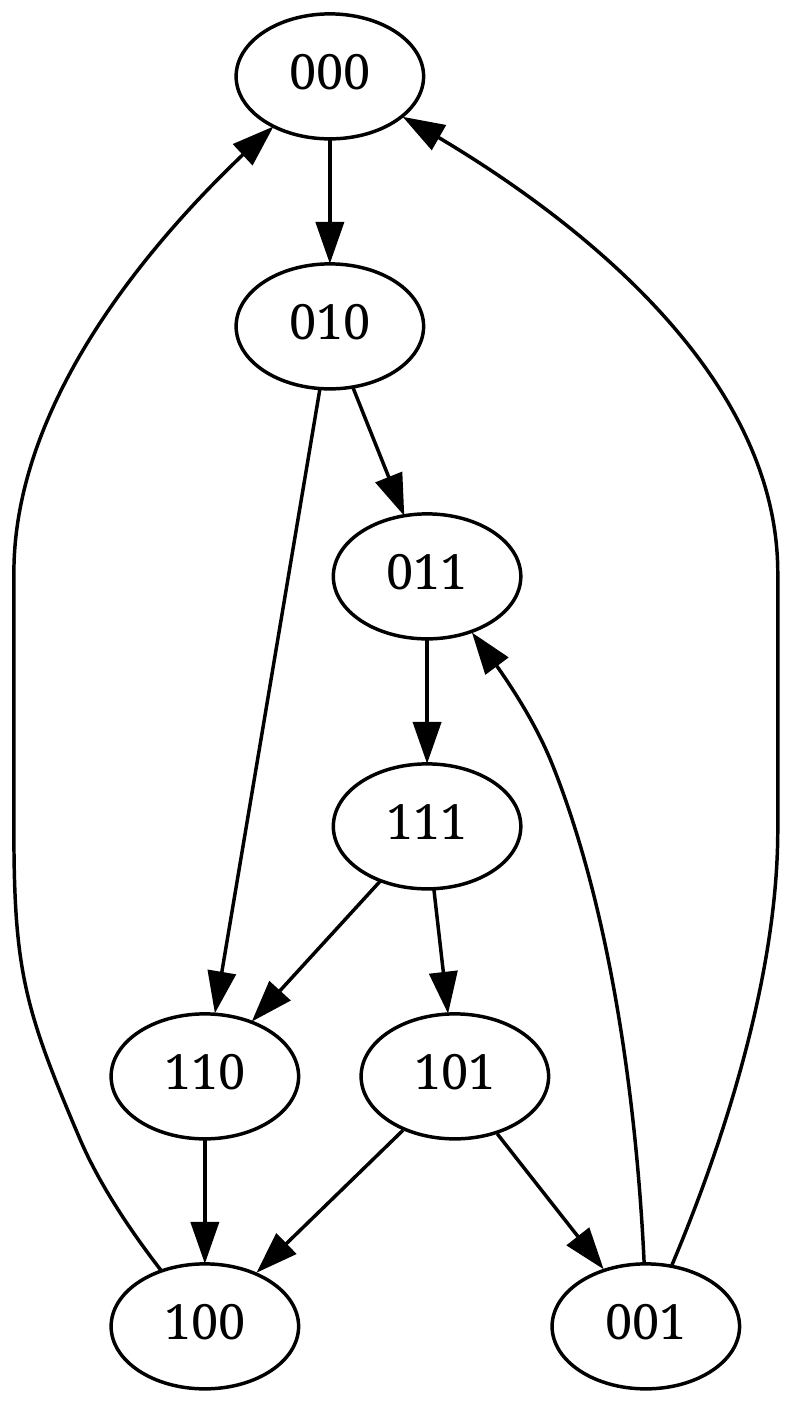}\caption{Fully-asynchronous dynamics of the example BN for P1}\label{fig:paper_files/paper_21_1.pdf}
    \end{figure}

    \noindent
    The resulting graphics is reproduced in Figure~\ref{fig:paper_files/paper_21_1.pdf}.

Recall that the fixed points are identical in asynchronous and MP. We
use \href{https://github.com/bnediction/mpbn}{\texttt{mpbn}} to analyze
the dynamical properties with the MP update mode:
\begin{small}
\begin{Verbatim}[commandchars=\\\{\}]
{\color{incolor}In [{\color{incolor}7}]:} \PY{n}{mf} \PY{o}{=} \PY{n}{mpbn}\PY{o}{.}\PY{n}{MPBooleanNetwork}\PY{p}{(}\PY{n}{f}\PY{p}{)}
        \PY{n+nb}{list}\PY{p}{(}\PY{n}{mf}\PY{o}{.}\PY{n}{fixedpoints}\PY{p}{(}\PY{p}{)}\PY{p}{)}
\end{Verbatim}
\end{small}

\begin{small}
\begin{Verbatim}[commandchars=\\\{\}]
{\color{outcolor}Out[{\color{outcolor}7}]:} []
\end{Verbatim}
\end{small}
\begin{small}
\begin{Verbatim}[commandchars=\\\{\}]
{\color{incolor}In [{\color{incolor}8}]:} \PY{n+nb}{list}\PY{p}{(}\PY{n}{mf}\PY{o}{.}\PY{n}{attractors}\PY{p}{(}\PY{p}{)}\PY{p}{)}
\end{Verbatim}
\end{small}

\begin{small}
\begin{Verbatim}[commandchars=\\\{\}]
{\color{outcolor}Out[{\color{outcolor}8}]:} [\{'A': '*', 'B': '*', 'C': '*'\}]
\end{Verbatim}
\end{small}
Indeed, the network has no fixed points, and its attractor is the full
hypercube of dimension 3.

Using the \texttt{marker\_reprogramming\_fixpoints} snippet defined
above, we identify all perturbations of at most 2 components which
ensure that (1) all the fixed points have \texttt{C} active, and (2) at
least one fixed point exists:
\begin{small}
\begin{Verbatim}[commandchars=\\\{\}]
{\color{incolor}In [{\color{incolor}9}]:} \PY{n+nb}{list}\PY{p}{(}\PY{n}{marker\PYZus{}reprogramming\PYZus{}fixpoints}\PY{p}{(}\PY{n}{f}\PY{p}{,} \PY{p}{\PYZob{}}\PY{l+s+s2}{\PYZdq{}}\PY{l+s+s2}{C}\PY{l+s+s2}{\PYZdq{}}\PY{p}{:} \PY{l+m+mi}{1}\PY{p}{\PYZcb{}}\PY{p}{,} \PY{l+m+mi}{2}\PY{p}{)}\PY{p}{)}
\end{Verbatim}
\end{small}

\begin{small}
\begin{Verbatim}[commandchars=\\\{\}]
{\color{outcolor}Out[{\color{outcolor}9}]:} [\{'A': 0\}, \{'C': 1, 'B': 0\}, \{'A': 1, 'C': 1\}, \{'B': 1, 'C': 1\}]
\end{Verbatim}
\end{small}
Indeed, fixing \texttt{A} to 0 breaks the negative feedback, and make
\texttt{B} converge to 1. There, \texttt{C} converges to state 1. Then,
remark that fixing \texttt{C} to 1 is not enough to fulfill the
property, as \texttt{A} and \texttt{B} still oscillate. Thus, one of
these 2 must be fixed as well, to any value. The solution
\texttt{\{\textquotesingle{}A\textquotesingle{}:\ 0,\ \textquotesingle{}C\textquotesingle{}:\ 1\}}
is not returned as
\texttt{\{\textquotesingle{}A\textquotesingle{}:\ 0\}} is sufficient to
acquire the desired dynamical property.

In our \emph{BoNesis} code snippet defined above, by default we ensure
that the perturbed BN possesses at least one fixed point. When relaxing
this constraint, we obtain that the empty perturbation is the (unique)
minimal solution, as \texttt{f} has no fixed point.
\begin{small}
\begin{Verbatim}[commandchars=\\\{\}]
{\color{incolor}In [{\color{incolor}10}]:} \PY{n+nb}{list}\PY{p}{(}\PY{n}{marker\PYZus{}reprogramming\PYZus{}fixpoints}\PY{p}{(}\PY{n}{f}\PY{p}{,} \PY{p}{\PYZob{}}\PY{l+s+s2}{\PYZdq{}}\PY{l+s+s2}{C}\PY{l+s+s2}{\PYZdq{}}\PY{p}{:} \PY{l+m+mi}{1}\PY{p}{\PYZcb{}}\PY{p}{,} \PY{l+m+mi}{2}\PY{p}{,} \PY{n}{ensure\PYZus{}exists}\PY{o}{=}\PY{k+kc}{False}\PY{p}{)}\PY{p}{)}
\end{Verbatim}
\end{small}

\begin{small}
\begin{Verbatim}[commandchars=\\\{\}]
{\color{outcolor}Out[{\color{outcolor}10}]:} [\{\}]
\end{Verbatim}
\end{small}
In the following, we demonstrate how to perform the same computation
with the command line. By default, the reprogramming of fixed points
adds the constraint that at least one fixed point must exist.
\begin{small}
\begin{Verbatim}[commandchars=\\\{\}]
{\color{incolor}In [{\color{incolor}11}]:} \PY{k}{with} \PY{n+nb}{open}\PY{p}{(}\PY{l+s+s2}{\PYZdq{}}\PY{l+s+s2}{example1.bnet}\PY{l+s+s2}{\PYZdq{}}\PY{p}{,} \PY{l+s+s2}{\PYZdq{}}\PY{l+s+s2}{w}\PY{l+s+s2}{\PYZdq{}}\PY{p}{)} \PY{k}{as} \PY{n}{fp}\PY{p}{:}
             \PY{n}{fp}\PY{o}{.}\PY{n}{write}\PY{p}{(}\PY{n}{f}\PY{o}{.}\PY{n}{source}\PY{p}{(}\PY{p}{)}\PY{p}{)}
         \PY{o}{\PYZpc{}}\PY{k}{cat} example1.bnet
\end{Verbatim}
\end{small}

    \begin{Verbatim}[commandchars=\\\{\}]
A, B
B, !A
C, !A\&B

    \end{Verbatim}

\begin{small}
\begin{Verbatim}[commandchars=\\\{\}]
{\color{incolor}In [{\color{incolor}12}]:} \PY{o}{!}bonesis\PYZhy{}reprogramming\PY{+w}{ }example1.bnet\PY{+w}{ }\PY{l+s+s1}{\PYZsq{}\PYZob{}\PYZdq{}C\PYZdq{}: 1\PYZcb{}\PYZsq{}}\PY{+w}{ }\PY{l+m}{2}\PY{+w}{ }\PYZhy{}\PYZhy{}fixpoints
\end{Verbatim}
\end{small}

    \begin{Verbatim}[commandchars=\\\{\}]
\{'A': 0\}
\{'C': 1, 'B': 0\}
\{'A': 1, 'C': 1\}
\{'B': 1, 'C': 1\}

    \end{Verbatim}

Adding the option \texttt{-\/-allow-no-fixpoint} would return an empty
perturbation as unique minimal solution.
\hypertarget{source-marker-reprogramming-of-fixed-points-p2}{%
\subsubsection{Source-marker reprogramming of fixed points
(P2)}\label{source-marker-reprogramming-of-fixed-points-p2}}

Given an initial configuration \(z\), we identify the perturbations
\(P\) of at most \(k\) components so that all the fixed points of
\(f/P\) that are reachable from \(z\) in \(f/P\) match with the given
marker \(M\). The associated decision problem can be expressed as the
following \(\exists\forall\)-expression, hence being at most in
\(\Sigma_2^{\mathrm P}\):

\begin{equation}
\exists P\in\mathbb M^k, \forall x\in\mathbb B^n, ((f/P)(x)=x \wedge \operatorname{reach}_P(z,x))\implies x\models M
\end{equation}

``There exists a perturbation \(P\) such that for any configuration
\(x\in\mathbb B^n\), if \(x\) is a fixed point of the perturbed BN
\((f/P)\), and \(x\) is reachable from \(z\) in \((f/P)\), then \(x\)
must match with \(M\)''.

As explained in the Method section, the reachability property boils down
to computing the smallest trap space containing \(z\): if it contains
the fixed point \(x\), then \(x\) is reachable from \(z\) with the MP
update mode.

\begin{equation}
\exists P\in\mathbb M^k, \forall x\in\mathbb B^n, ((f/P)(x)=x \wedge 
x\in\operatorname{TS}_P(z))\implies x\models M\enspace.
\end{equation}

As with the previous case, in practice we may also want that there
exists at least one fixed point reachable from \(z\).

With the \emph{BoNesis} Python interface, this reprogramming property is
declared as follows, where \texttt{f} is a BN, \texttt{z} the initial
configuration (Python dictionary), \texttt{M} the marker, and \texttt{k}
the maximum number of components that can be perturbed (at most \(n\)):
\begin{small}
\begin{Verbatim}[commandchars=\\\{\}]
{\color{incolor}In [{\color{incolor}13}]:} \PY{k}{def} \PY{n+nf}{source\PYZus{}marker\PYZus{}reprogramming\PYZus{}fixpoints}\PY{p}{(}\PY{n}{f}\PY{p}{:} \PY{n}{BooleanNetwork}\PY{p}{,}
                                                   \PY{n}{z}\PY{p}{:} \PY{n+nb}{dict}\PY{p}{[}\PY{n+nb}{str}\PY{p}{,}\PY{n+nb}{bool}\PY{p}{]}\PY{p}{,}
                                                   \PY{n}{M}\PY{p}{:} \PY{n+nb}{dict}\PY{p}{[}\PY{n+nb}{str}\PY{p}{,}\PY{n+nb}{bool}\PY{p}{]}\PY{p}{,}
                                                   \PY{n}{k}\PY{p}{:} \PY{n+nb}{int}\PY{p}{)}\PY{p}{:}
             \PY{c+c1}{\PYZsh{} f: Boolean network; z: initial configuration;}
             \PY{c+c1}{\PYZsh{} M: marker; k: maximum number of components to perturb}
             \PY{n}{bo} \PY{o}{=} \PY{n}{bonesis}\PY{o}{.}\PY{n}{BoNesis}\PY{p}{(}\PY{n}{f}\PY{p}{)}
             \PY{n}{P} \PY{o}{=} \PY{n}{bo}\PY{o}{.}\PY{n}{Some}\PY{p}{(}\PY{n}{max\PYZus{}size}\PY{o}{=}\PY{n}{k}\PY{p}{)}  \PY{c+c1}{\PYZsh{} perturbation to identify}
             \PY{k}{with} \PY{n}{bo}\PY{o}{.}\PY{n}{mutant}\PY{p}{(}\PY{n}{P}\PY{p}{)}\PY{p}{:}
                 \PY{c+c1}{\PYZsh{} all the fixed points reachable from z match with M}
                 \PY{o}{\PYZti{}}\PY{n}{bo}\PY{o}{.}\PY{n}{obs}\PY{p}{(}\PY{n}{z}\PY{p}{)} \PY{o}{\PYZgt{}\PYZgt{}} \PY{l+s+s2}{\PYZdq{}}\PY{l+s+s2}{fixpoints}\PY{l+s+s2}{\PYZdq{}} \PY{o}{\PYZca{}} \PY{p}{\PYZob{}}\PY{n}{bo}\PY{o}{.}\PY{n}{obs}\PY{p}{(}\PY{n}{M}\PY{p}{)}\PY{p}{\PYZcb{}}
                 \PY{c+c1}{\PYZsh{} at least one fixed point matching with M is reachable from z}
                 \PY{o}{\PYZti{}}\PY{n}{bo}\PY{o}{.}\PY{n}{obs}\PY{p}{(}\PY{n}{z}\PY{p}{)} \PY{o}{\PYZgt{}}\PY{o}{=} \PY{n}{bo}\PY{o}{.}\PY{n}{fixed}\PY{p}{(}\PY{o}{\PYZti{}}\PY{n}{bo}\PY{o}{.}\PY{n}{obs}\PY{p}{(}\PY{n}{M}\PY{p}{)}\PY{p}{)}
             \PY{k}{return} \PY{n}{P}\PY{o}{.}\PY{n}{assignments}\PY{p}{(}\PY{p}{)}
\end{Verbatim}
\end{small}

Compared to the previous code snippet, this function relies on specific
operators to restrict the properties to the fixed point reachable from
\texttt{z}. The instruction \texttt{\textasciitilde{}bo.obs(z)} refers
to a specific configuration matching with \texttt{z};
\texttt{\textasciitilde{}bo.obs(z)\ \textgreater{}\textgreater{}\ "fixpoints"\ \^{}\ \{bo.obs(M)\}}
specifies that all the fixed points reachable from such configuration
have to match with at least one configuration given in the set
\texttt{\{bo.obs(M)\}}, i.e., \texttt{M} in this case. This property is
satisfied whenever no fixed point are reachable. Thus, the next line
ensures that at least one fixed point is reachable from the
configuration associated with \texttt{z}:
\texttt{bo.fixed(\textasciitilde{}bo.obs(M))} refers to one
configuration which is a fixed point (in the perturbed BN), and which
matches with \texttt{M}. Then, the binary operator
\texttt{\textgreater{}=} declares the existence of a trajectory from its
left to its right configuration.

Notice that with this formulation, in the case whenever \texttt{z} is
only partially defined (some components are not associated to a Boolean
value), a perturbation is returned as long as there exists at least one
fully-defined configuration matching with \(z\) which fulfil the
specified dynamical properties.

The corresponding command line is of the form

\begin{Shaded}
\begin{Highlighting}[]
\ExtensionTok{bonesis{-}reprogramming}\NormalTok{ model.bnet M k }\AttributeTok{{-}{-}fixpoints} \AttributeTok{{-}{-}reachable{-}from}\NormalTok{ z}
\end{Highlighting}
\end{Shaded}

where \texttt{model.bnet} is a BN encoded in BooleanNet format,
\texttt{M} specifies the marker as a JSON map, \texttt{k} is the maximum
number of perturbations, and \texttt{z} is the initial configuration as
a JSON map. The existence of at least one fixed point can be lifted with
the option \texttt{-\/-allow-no-fixpoint}.
\paragraph{Example}

Let us consider the following toy BN with two positive feedback cycles:
\begin{small}
\begin{Verbatim}[commandchars=\\\{\}]
{\color{incolor}In [{\color{incolor}14}]:} \PY{n}{f} \PY{o}{=} \PY{n}{BooleanNetwork}\PY{p}{(}\PY{p}{\PYZob{}}
             \PY{l+s+s2}{\PYZdq{}}\PY{l+s+s2}{A}\PY{l+s+s2}{\PYZdq{}}\PY{p}{:} \PY{l+s+s2}{\PYZdq{}}\PY{l+s+s2}{B}\PY{l+s+s2}{\PYZdq{}}\PY{p}{,}
             \PY{l+s+s2}{\PYZdq{}}\PY{l+s+s2}{B}\PY{l+s+s2}{\PYZdq{}}\PY{p}{:} \PY{l+s+s2}{\PYZdq{}}\PY{l+s+s2}{A}\PY{l+s+s2}{\PYZdq{}}\PY{p}{,}
             \PY{l+s+s2}{\PYZdq{}}\PY{l+s+s2}{C}\PY{l+s+s2}{\PYZdq{}}\PY{p}{:} \PY{l+s+s2}{\PYZdq{}}\PY{l+s+s2}{!D \PYZam{} (A|B)}\PY{l+s+s2}{\PYZdq{}}\PY{p}{,}
             \PY{l+s+s2}{\PYZdq{}}\PY{l+s+s2}{D}\PY{l+s+s2}{\PYZdq{}}\PY{p}{:} \PY{l+s+s2}{\PYZdq{}}\PY{l+s+s2}{!C}\PY{l+s+s2}{\PYZdq{}}
         \PY{p}{\PYZcb{}}\PY{p}{)}
         \PY{n}{f}\PY{o}{.}\PY{n}{influence\PYZus{}graph}\PY{p}{(}\PY{p}{)}
\end{Verbatim}
\end{small}

\begin{figure}[tbp]
    \centering
    \includegraphics[scale=0.7]{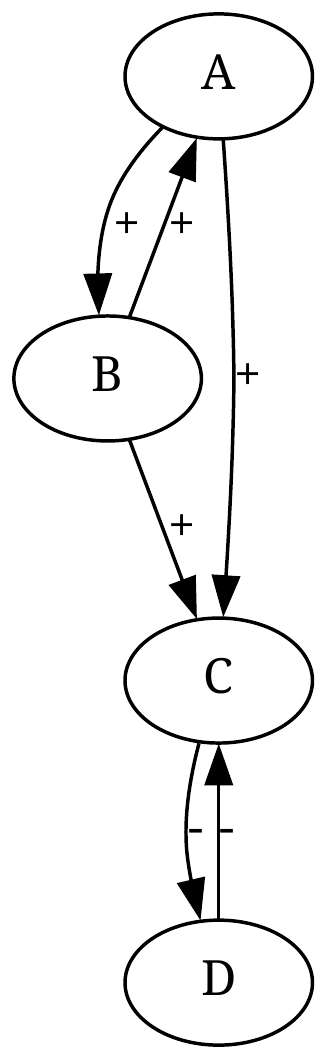}\caption{Influence graph of the example BN for P2}\label{fig:paper_files/paper_37_1.pdf}
    \end{figure}

    \noindent
    The resulting graphics is reproduced in Figure~\ref{fig:paper_files/paper_37_1.pdf}.

This BN has 3 fixed points, 2 of which are reachable from the
configuration where \texttt{A} and \texttt{B} are active, and \texttt{C}
and \texttt{D} inactive:
\begin{small}
\begin{Verbatim}[commandchars=\\\{\}]
{\color{incolor}In [{\color{incolor}15}]:} \PY{n}{z} \PY{o}{=} \PY{p}{\PYZob{}}\PY{l+s+s2}{\PYZdq{}}\PY{l+s+s2}{A}\PY{l+s+s2}{\PYZdq{}}\PY{p}{:} \PY{l+m+mi}{1}\PY{p}{,} \PY{l+s+s2}{\PYZdq{}}\PY{l+s+s2}{B}\PY{l+s+s2}{\PYZdq{}}\PY{p}{:} \PY{l+m+mi}{1}\PY{p}{,} \PY{l+s+s2}{\PYZdq{}}\PY{l+s+s2}{C}\PY{l+s+s2}{\PYZdq{}}\PY{p}{:} \PY{l+m+mi}{0}\PY{p}{,} \PY{l+s+s2}{\PYZdq{}}\PY{l+s+s2}{D}\PY{l+s+s2}{\PYZdq{}}\PY{p}{:} \PY{l+m+mi}{0}\PY{p}{\PYZcb{}}
         \PY{n}{f}\PY{o}{.}\PY{n}{dynamics}\PY{p}{(}\PY{l+s+s2}{\PYZdq{}}\PY{l+s+s2}{asynchronous}\PY{l+s+s2}{\PYZdq{}}\PY{p}{,} \PY{n}{init}\PY{o}{=}\PY{n}{z}\PY{p}{)}
\end{Verbatim}
\end{small}

\begin{figure}[tbp]
    \centering
    \includegraphics[scale=0.7]{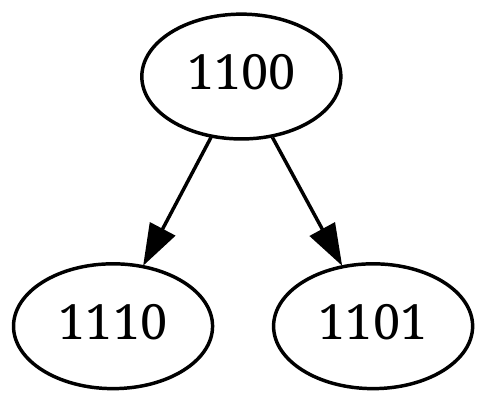}\caption{Fully-asynchronous dynamics of the example BN for P2 from the configuration $1100$}\label{fig:paper_files/paper_39_1.pdf}
    \end{figure}

    \noindent
    The resulting graphics is reproduced in Figure~\ref{fig:paper_files/paper_39_1.pdf}.

\begin{small}
\begin{Verbatim}[commandchars=\\\{\}]
{\color{incolor}In [{\color{incolor}16}]:} \PY{n+nb}{list}\PY{p}{(}\PY{n}{mpbn}\PY{o}{.}\PY{n}{MPBooleanNetwork}\PY{p}{(}\PY{n}{f}\PY{p}{)}\PY{o}{.}\PY{n}{fixedpoints}\PY{p}{(}\PY{p}{)}\PY{p}{)}
\end{Verbatim}
\end{small}

\begin{small}
\begin{Verbatim}[commandchars=\\\{\}]
{\color{outcolor}Out[{\color{outcolor}16}]:} [\{'A': 0, 'B': 0, 'C': 0, 'D': 1\},
          \{'A': 1, 'B': 1, 'C': 0, 'D': 1\},
          \{'A': 1, 'B': 1, 'C': 1, 'D': 0\}]
\end{Verbatim}
\end{small}
\begin{small}
\begin{Verbatim}[commandchars=\\\{\}]
{\color{incolor}In [{\color{incolor}17}]:} \PY{n+nb}{list}\PY{p}{(}\PY{n}{mpbn}\PY{o}{.}\PY{n}{MPBooleanNetwork}\PY{p}{(}\PY{n}{f}\PY{p}{)}\PY{o}{.}\PY{n}{fixedpoints}\PY{p}{(}\PY{n}{reachable\PYZus{}from}\PY{o}{=}\PY{n}{z}\PY{p}{)}\PY{p}{)}
\end{Verbatim}
\end{small}

\begin{small}
\begin{Verbatim}[commandchars=\\\{\}]
{\color{outcolor}Out[{\color{outcolor}17}]:} [\{'A': 1, 'B': 1, 'C': 1, 'D': 0\}, \{'A': 1, 'B': 1, 'C': 0, 'D': 1\}]
\end{Verbatim}
\end{small}
Let us compare the results of the global marker-reprogramming of fixed
points (P1) with the source-marker reprogramming of fixed points (P2),
the objective being to have fixed points having \texttt{C} active. In
the first case, putting aside the perturbation of \texttt{C}, this
necessitates to act on either \texttt{A} or \texttt{B} to prevent the
existence of the fixed points where \texttt{A}, \texttt{B} and
\texttt{C} are inactive:
\begin{small}
\begin{Verbatim}[commandchars=\\\{\}]
{\color{incolor}In [{\color{incolor}18}]:} \PY{n+nb}{list}\PY{p}{(}\PY{n}{marker\PYZus{}reprogramming\PYZus{}fixpoints}\PY{p}{(}\PY{n}{f}\PY{p}{,} \PY{p}{\PYZob{}}\PY{l+s+s2}{\PYZdq{}}\PY{l+s+s2}{C}\PY{l+s+s2}{\PYZdq{}}\PY{p}{:} \PY{l+m+mi}{1}\PY{p}{\PYZcb{}}\PY{p}{,} \PY{l+m+mi}{2}\PY{p}{)}\PY{p}{)}
\end{Verbatim}
\end{small}

\begin{small}
\begin{Verbatim}[commandchars=\\\{\}]
{\color{outcolor}Out[{\color{outcolor}18}]:} [\{'A': 1, 'D': 0\}, \{'B': 1, 'D': 0\}, \{'C': 1\}]
\end{Verbatim}
\end{small}
Considering only the fixed points reachable from the configuration
\texttt{z}, there is no need to act on \texttt{A} or \texttt{B}:
\begin{small}
\begin{Verbatim}[commandchars=\\\{\}]
{\color{incolor}In [{\color{incolor}19}]:} \PY{n+nb}{list}\PY{p}{(}\PY{n}{source\PYZus{}marker\PYZus{}reprogramming\PYZus{}fixpoints}\PY{p}{(}\PY{n}{f}\PY{p}{,} \PY{n}{z}\PY{p}{,} \PY{p}{\PYZob{}}\PY{l+s+s2}{\PYZdq{}}\PY{l+s+s2}{C}\PY{l+s+s2}{\PYZdq{}}\PY{p}{:} \PY{l+m+mi}{1}\PY{p}{\PYZcb{}}\PY{p}{,} \PY{l+m+mi}{2}\PY{p}{)}\PY{p}{)}
\end{Verbatim}
\end{small}

\begin{small}
\begin{Verbatim}[commandchars=\\\{\}]
{\color{outcolor}Out[{\color{outcolor}19}]:} [\{'D': 0\}, \{'C': 1\}]
\end{Verbatim}
\end{small}
The command line program \texttt{bonesis-reprogramming} can perform P2
by specifying the \texttt{-\/-rechable-from} option giving the initial
configuration in JSON format:
\begin{small}
\begin{Verbatim}[commandchars=\\\{\}]
{\color{incolor}In [{\color{incolor}20}]:} \PY{k}{with} \PY{n+nb}{open}\PY{p}{(}\PY{l+s+s2}{\PYZdq{}}\PY{l+s+s2}{example2.bnet}\PY{l+s+s2}{\PYZdq{}}\PY{p}{,} \PY{l+s+s2}{\PYZdq{}}\PY{l+s+s2}{w}\PY{l+s+s2}{\PYZdq{}}\PY{p}{)} \PY{k}{as} \PY{n}{fp}\PY{p}{:}
             \PY{n}{fp}\PY{o}{.}\PY{n}{write}\PY{p}{(}\PY{n}{f}\PY{o}{.}\PY{n}{source}\PY{p}{(}\PY{p}{)}\PY{p}{)}
\end{Verbatim}
\end{small}

\begin{small}
\begin{Verbatim}[commandchars=\\\{\}]
{\color{incolor}In [{\color{incolor}21}]:} \PY{o}{!}bonesis\PYZhy{}reprogramming\PY{+w}{ }example2.bnet\PY{+w}{ }\PY{l+s+s1}{\PYZsq{}\PYZob{}\PYZdq{}C\PYZdq{}: 1\PYZcb{}\PYZsq{}}\PY{+w}{ }\PY{l+m}{2}\PY{+w}{ }\PYZhy{}\PYZhy{}fixpoints\PY{+w}{ }\PY{err}{\PYZbs{}}
             \PY{o}{\PYZhy{}}\PY{o}{\PYZhy{}}\PY{n}{reachable}\PY{o}{\PYZhy{}}\PY{k+kn}{from} \PY{l+s+s1}{\PYZsq{}}\PY{l+s+s1}{\PYZob{}}\PY{l+s+s1}{\PYZdq{}}\PY{l+s+s1}{A}\PY{l+s+s1}{\PYZdq{}}\PY{l+s+s1}{: 1, }\PY{l+s+s1}{\PYZdq{}}\PY{l+s+s1}{B}\PY{l+s+s1}{\PYZdq{}}\PY{l+s+s1}{: 1, }\PY{l+s+s1}{\PYZdq{}}\PY{l+s+s1}{C}\PY{l+s+s1}{\PYZdq{}}\PY{l+s+s1}{: 0, }\PY{l+s+s1}{\PYZdq{}}\PY{l+s+s1}{D}\PY{l+s+s1}{\PYZdq{}}\PY{l+s+s1}{: 0\PYZcb{}}\PY{l+s+s1}{\PYZsq{}}
\end{Verbatim}
\end{small}

    \begin{Verbatim}[commandchars=\\\{\}]
\{'D': 0\}
\{'C': 1\}

    \end{Verbatim}

\hypertarget{marker-reprogramming-of-attractors-p3}{%
\subsubsection{Marker reprogramming of attractors
(P3)}\label{marker-reprogramming-of-attractors-p3}}

We identify the perturbations \(P\) of at most \(k\) components so that
the configurations of the all the attractors of \(f/P\) match with the
given marker \(M\) (i.e., in each attractor, the specified markers
cannot oscillate). The associated decision problem can be expressed as
follows:

\begin{equation}
\exists P\in\mathbb M^k, \forall x\in\mathbb B^n, \operatorname{IN-ATTRACTOR}_P(x) \implies x\models M
\end{equation}

(``There exists a perturbation \(P\) of at most \(k\) components, such
that for all configurations \(x\), if \(x\) belongs to an attractor of
the perturbed BN \(f/P\), then \(x\) matches with the specified markers
\(M\)'')

By restricting the range of the universal part of the equation to the
configurations which do not match with the marker \(M\), we obtain:
\begin{equation}
\exists P\in\mathbb M^k, \forall x\in\mathbb B^n: x\not\models M, \neg\operatorname{IN-ATTRACTOR}_P(x)
\end{equation}

The \(\operatorname{IN-ATTRACTOR}\) property being itself a quantified
Boolean expression, we obtain the following
\(\exists\forall\exists\)-expression:

\begin{equation}
\exists P\in\mathbb M^k, \forall x\in\mathbb B^n: x\not\models M, \exists y\in\mathbb B^n,
   y\in \operatorname{TS}_P(x), \operatorname{TS}_P(y) \neq \operatorname{TS}_P(x)
\end{equation}

The problem of satisfiability of this quantified Boolean expression is
beyond the expressiveness power of ASP which is limited to
\(\exists\forall\)-expressions. Nevertheless, we can approach this
problem by its complementary: the existence of perturbations of size
\(k\) such that at least one configuration belonging to an attractor
does \emph{not} match with the marker \(M\). This complementary problem
can be expressed with this following expression \begin{equation}
\exists P\in\mathbb M^k, \exists x\in\mathbb B^n,x\not\models M \wedge \operatorname{IN-ATTRACTOR}_P(x)
\end{equation} which is an \(\exists\forall\)-expression:
\begin{equation}
\exists P\in\mathbb M^k, \exists x\in\mathbb B^n, x\not\models M\wedge \forall y\in\mathbb B^n, y\in \operatorname{TS}_P(x) \implies \operatorname{TS}_P(y) \neq \operatorname{TS}_P(x)
\enspace.
\end{equation}

Because the domain of candidate perturbations \(\mathbb M^{\leq k}\) is
finite, one can first resolve the complementary problem, giving all bad
perturbations, and returns its complement.

Notice that this approach is highly combinatorics, and is likely limited
to identifying perturbations of small size. However, to our knowledge,
this is the first implemented method which addresses the complement
reprogramming of MP attractors, i.e., of the minimal trap spaces of the
BN.

With the \emph{BoNesis} Python interface, this reprogramming property is
declared as follows, where \texttt{f} is a BN, \texttt{M} the marker,
and \texttt{k} the maximum number of components that can be perturbed
(at most \(n\)):
\begin{small}
\begin{Verbatim}[commandchars=\\\{\}]
{\color{incolor}In [{\color{incolor}22}]:} \PY{k}{def} \PY{n+nf}{marker\PYZus{}reprogramming}\PY{p}{(}\PY{n}{f}\PY{p}{:} \PY{n}{BooleanNetwork}\PY{p}{,}
                                  \PY{n}{M}\PY{p}{:} \PY{n+nb}{dict}\PY{p}{[}\PY{n+nb}{str}\PY{p}{,}\PY{n+nb}{bool}\PY{p}{]}\PY{p}{,}
                                  \PY{n}{k}\PY{p}{:} \PY{n+nb}{int}\PY{p}{)}\PY{p}{:}
             \PY{n}{bo} \PY{o}{=} \PY{n}{bonesis}\PY{o}{.}\PY{n}{BoNesis}\PY{p}{(}\PY{n}{f}\PY{p}{)}
             \PY{n}{coP} \PY{o}{=} \PY{n}{bo}\PY{o}{.}\PY{n}{Some}\PY{p}{(}\PY{n}{max\PYZus{}size}\PY{o}{=}\PY{n}{k}\PY{p}{)}
             \PY{k}{with} \PY{n}{bo}\PY{o}{.}\PY{n}{mutant}\PY{p}{(}\PY{n}{coP}\PY{p}{)}\PY{p}{:}
                 \PY{n}{x} \PY{o}{=} \PY{n}{bo}\PY{o}{.}\PY{n}{cfg}\PY{p}{(}\PY{p}{)}
                 \PY{n}{bo}\PY{o}{.}\PY{n}{in\PYZus{}attractor}\PY{p}{(}\PY{n}{x}\PY{p}{)}
                 \PY{n}{x} \PY{o}{!=} \PY{n}{bo}\PY{o}{.}\PY{n}{obs}\PY{p}{(}\PY{n}{M}\PY{p}{)}
             \PY{k}{return} \PY{n}{coP}\PY{o}{.}\PY{n}{complementary\PYZus{}assignments}\PY{p}{(}\PY{p}{)}
\end{Verbatim}
\end{small}

The idea of the above code snippet is to declare the property of being a
bad perturbation (\texttt{coP}). In the context of this perturbation, we
declare with \texttt{x=bo.cfg()} a configuration. Then,
\texttt{bo.in\_attractor(x)} imposes that \texttt{x} belongs to an
attractor (minimal trap space) of the perturbed BN; and the
\texttt{x\ !=\ bo.obs(M)} instruction adds the constraint that
\texttt{x} must not match with \texttt{M}.

The \texttt{.complementary\_assignments()} method performs the
computation of the complement of solutions. It solves the above problem
with perturbation sizes ranging from 0 to \texttt{k}, and returns
complement minimal perturbations only.

The corresponding command line is of the form

\begin{Shaded}
\begin{Highlighting}[]
\ExtensionTok{bonesis{-}reprogramming}\NormalTok{ model.bnet M k}
\end{Highlighting}
\end{Shaded}

where \texttt{model.bnet} is a BN encoded in BooleanNet format,
\texttt{M} specifies the marker as a JSON map, \texttt{k} is the maximum
number of perturbations.
\paragraph{Example}

Let us consider the following BN:
\begin{small}
\begin{Verbatim}[commandchars=\\\{\}]
{\color{incolor}In [{\color{incolor}23}]:} \PY{n}{f} \PY{o}{=} \PY{n}{mpbn}\PY{o}{.}\PY{n}{MPBooleanNetwork}\PY{p}{(}\PY{p}{\PYZob{}}
             \PY{l+s+s2}{\PYZdq{}}\PY{l+s+s2}{A}\PY{l+s+s2}{\PYZdq{}}\PY{p}{:} \PY{l+s+s2}{\PYZdq{}}\PY{l+s+s2}{!B}\PY{l+s+s2}{\PYZdq{}}\PY{p}{,}
             \PY{l+s+s2}{\PYZdq{}}\PY{l+s+s2}{B}\PY{l+s+s2}{\PYZdq{}}\PY{p}{:} \PY{l+s+s2}{\PYZdq{}}\PY{l+s+s2}{!A}\PY{l+s+s2}{\PYZdq{}}\PY{p}{,}
             \PY{l+s+s2}{\PYZdq{}}\PY{l+s+s2}{C}\PY{l+s+s2}{\PYZdq{}}\PY{p}{:} \PY{l+s+s2}{\PYZdq{}}\PY{l+s+s2}{A \PYZam{} !B \PYZam{} !D}\PY{l+s+s2}{\PYZdq{}}\PY{p}{,}
             \PY{l+s+s2}{\PYZdq{}}\PY{l+s+s2}{D}\PY{l+s+s2}{\PYZdq{}}\PY{p}{:} \PY{l+s+s2}{\PYZdq{}}\PY{l+s+s2}{C | E}\PY{l+s+s2}{\PYZdq{}}\PY{p}{,}
             \PY{l+s+s2}{\PYZdq{}}\PY{l+s+s2}{E}\PY{l+s+s2}{\PYZdq{}}\PY{p}{:} \PY{l+s+s2}{\PYZdq{}}\PY{l+s+s2}{!C \PYZam{} !E}\PY{l+s+s2}{\PYZdq{}}\PY{p}{,}
         \PY{p}{\PYZcb{}}\PY{p}{)}
         \PY{n}{f}\PY{o}{.}\PY{n}{influence\PYZus{}graph}\PY{p}{(}\PY{p}{)}
\end{Verbatim}
\end{small}

\begin{figure}[tbp]
    \centering
    \includegraphics[scale=0.7]{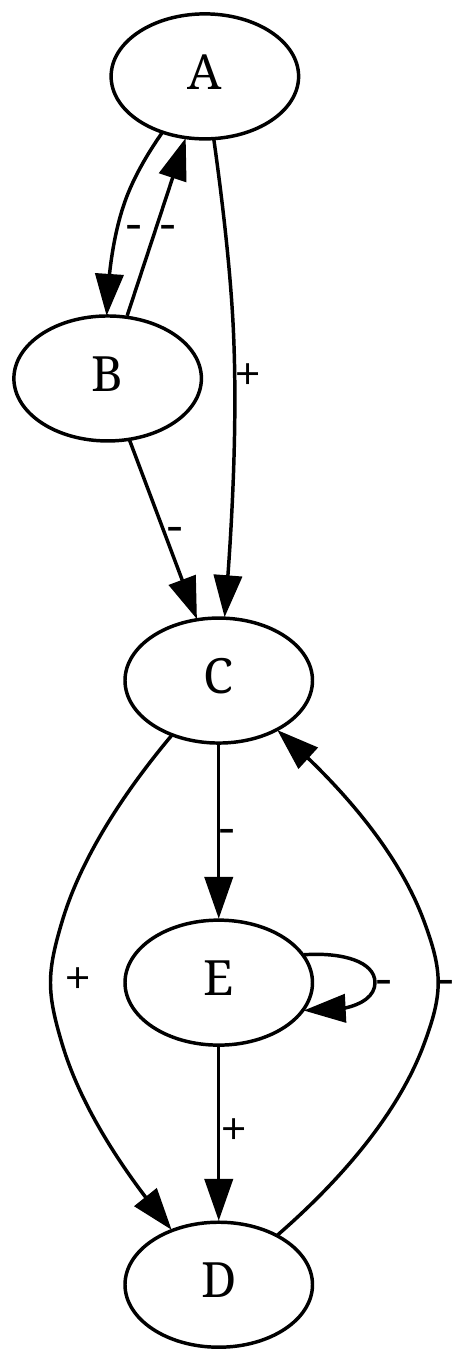}\caption{Influence graph of the example BN for P3}\label{fig:paper_files/paper_53_1.pdf}
    \end{figure}

    \noindent
    The resulting graphics is reproduced in Figure~\ref{fig:paper_files/paper_53_1.pdf}.

Essentially, \texttt{A} and \texttt{B} always stabilize to opposite
states. Whenever \texttt{A} is active (and \texttt{B} inactive) then
\texttt{C} will oscillate, otherwise it stabilizes to 0. In each case
\texttt{D} and \texttt{E} oscillate. This lead to the following MP
attractors:
\begin{small}
\begin{Verbatim}[commandchars=\\\{\}]
{\color{incolor}In [{\color{incolor}24}]:} \PY{n}{tabulate}\PY{p}{(}\PY{n+nb}{list}\PY{p}{(}\PY{n}{f}\PY{o}{.}\PY{n}{attractors}\PY{p}{(}\PY{p}{)}\PY{p}{)}\PY{p}{)}
\end{Verbatim}
\end{small}

\begin{small}
\begin{Verbatim}[commandchars=\\\{\}]
{\color{outcolor}Out[{\color{outcolor}24}]:}    A  B  C  D  E
         0  0  1  0  *  *
         1  1  0  *  *  *
\end{Verbatim}
\end{small}
Let us say that our objective is to reprogram the BN such that all the
attractors of the component \texttt{C} fixed to 1. The reprogramming of
fixed points (P1) gives the following solutions:
\begin{small}
\begin{Verbatim}[commandchars=\\\{\}]
{\color{incolor}In [{\color{incolor}25}]:} \PY{n+nb}{list}\PY{p}{(}\PY{n}{marker\PYZus{}reprogramming\PYZus{}fixpoints}\PY{p}{(}\PY{n}{f}\PY{p}{,} \PY{p}{\PYZob{}}\PY{l+s+s2}{\PYZdq{}}\PY{l+s+s2}{C}\PY{l+s+s2}{\PYZdq{}}\PY{p}{:} \PY{l+m+mi}{1}\PY{p}{\PYZcb{}}\PY{p}{,} \PY{l+m+mi}{3}\PY{p}{)}\PY{p}{)}
\end{Verbatim}
\end{small}

\begin{small}
\begin{Verbatim}[commandchars=\\\{\}]
{\color{outcolor}Out[{\color{outcolor}25}]:} [\{'D': 0\}, \{'C': 1\}]
\end{Verbatim}
\end{small}
Putting aside the trivial solution of perturbing \texttt{C}, let us
analyze the BN perturbed with the \texttt{D} forced to 0:
\begin{small}
\begin{Verbatim}[commandchars=\\\{\}]
{\color{incolor}In [{\color{incolor}26}]:} \PY{n}{pf} \PY{o}{=} \PY{n}{f}\PY{o}{.}\PY{n}{copy}\PY{p}{(}\PY{p}{)}
         \PY{n}{pf}\PY{p}{[}\PY{l+s+s2}{\PYZdq{}}\PY{l+s+s2}{D}\PY{l+s+s2}{\PYZdq{}}\PY{p}{]} \PY{o}{=} \PY{l+m+mi}{0}
         \PY{n}{tabulate}\PY{p}{(}\PY{n}{pf}\PY{o}{.}\PY{n}{attractors}\PY{p}{(}\PY{p}{)}\PY{p}{)}
\end{Verbatim}
\end{small}

\begin{small}
\begin{Verbatim}[commandchars=\\\{\}]
{\color{outcolor}Out[{\color{outcolor}26}]:}    A  B  C  D  E
         1  0  1  0  0  *
         0  1  0  1  0  0
\end{Verbatim}
\end{small}
The (only) fixed point of the network indeed has \texttt{C} active.
However, it possesses another (cyclic) attractor, where \texttt{C} is
inactive. This example points out that focusing on fixed point
reprogramming may lead to predicting perturbations which are not
sufficient to ensure that all the attractors show the desired marker.

The complete attractor reprogramming returns that the perturbation of
\texttt{D} must be coupled with a perturbation of \texttt{A} or
\texttt{B}, in this case to destroy the cyclic attractor.
\begin{small}
\begin{Verbatim}[commandchars=\\\{\}]
{\color{incolor}In [{\color{incolor}27}]:} \PY{n+nb}{list}\PY{p}{(}\PY{n}{marker\PYZus{}reprogramming}\PY{p}{(}\PY{n}{f}\PY{p}{,} \PY{p}{\PYZob{}}\PY{l+s+s2}{\PYZdq{}}\PY{l+s+s2}{C}\PY{l+s+s2}{\PYZdq{}}\PY{p}{:} \PY{l+m+mi}{1}\PY{p}{\PYZcb{}}\PY{p}{,} \PY{l+m+mi}{3}\PY{p}{)}\PY{p}{)}
\end{Verbatim}
\end{small}

\begin{small}
\begin{Verbatim}[commandchars=\\\{\}]
{\color{outcolor}Out[{\color{outcolor}27}]:} [\{'C': 1\}, \{'D': 0, 'B': 0\}, \{'D': 0, 'A': 1\}]
\end{Verbatim}
\end{small}
The same results can be obtained using the command line as follows.
\begin{small}
\begin{Verbatim}[commandchars=\\\{\}]
{\color{incolor}In [{\color{incolor}28}]:} \PY{k}{with} \PY{n+nb}{open}\PY{p}{(}\PY{l+s+s2}{\PYZdq{}}\PY{l+s+s2}{example3.bnet}\PY{l+s+s2}{\PYZdq{}}\PY{p}{,} \PY{l+s+s2}{\PYZdq{}}\PY{l+s+s2}{w}\PY{l+s+s2}{\PYZdq{}}\PY{p}{)} \PY{k}{as} \PY{n}{fp}\PY{p}{:}
             \PY{n}{fp}\PY{o}{.}\PY{n}{write}\PY{p}{(}\PY{n}{f}\PY{o}{.}\PY{n}{source}\PY{p}{(}\PY{p}{)}\PY{p}{)}
\end{Verbatim}
\end{small}

\begin{small}
\begin{Verbatim}[commandchars=\\\{\}]
{\color{incolor}In [{\color{incolor}29}]:} \PY{o}{!}bonesis\PYZhy{}reprogramming\PY{+w}{ }example3.bnet\PY{+w}{ }\PY{l+s+s1}{\PYZsq{}\PYZob{}\PYZdq{}C\PYZdq{}: 1\PYZcb{}\PYZsq{}}\PY{+w}{ }\PY{l+m}{3}
\end{Verbatim}
\end{small}

    \begin{Verbatim}[commandchars=\\\{\}]
\{'C': 1\}
\{'D': 0, 'B': 0\}
\{'D': 0, 'A': 1\}

    \end{Verbatim}

In other cases, the reprogramming of attractors may also lead to fewer
required perturbations than focusing on fixed points, provided we
enforce the existence of at least one fixed point. This can be
illustrated with the following example:
\begin{small}
\begin{Verbatim}[commandchars=\\\{\}]
{\color{incolor}In [{\color{incolor}30}]:} \PY{n}{g} \PY{o}{=} \PY{n}{mpbn}\PY{o}{.}\PY{n}{MPBooleanNetwork}\PY{p}{(}\PY{p}{\PYZob{}}
             \PY{l+s+s2}{\PYZdq{}}\PY{l+s+s2}{A}\PY{l+s+s2}{\PYZdq{}}\PY{p}{:} \PY{l+s+s2}{\PYZdq{}}\PY{l+s+s2}{!B}\PY{l+s+s2}{\PYZdq{}}\PY{p}{,}
             \PY{l+s+s2}{\PYZdq{}}\PY{l+s+s2}{B}\PY{l+s+s2}{\PYZdq{}}\PY{p}{:} \PY{l+s+s2}{\PYZdq{}}\PY{l+s+s2}{A}\PY{l+s+s2}{\PYZdq{}}\PY{p}{,}
             \PY{l+s+s2}{\PYZdq{}}\PY{l+s+s2}{C}\PY{l+s+s2}{\PYZdq{}}\PY{p}{:} \PY{l+s+s2}{\PYZdq{}}\PY{l+s+s2}{A \PYZam{} B}\PY{l+s+s2}{\PYZdq{}}\PY{p}{,}
             \PY{l+s+s2}{\PYZdq{}}\PY{l+s+s2}{D}\PY{l+s+s2}{\PYZdq{}}\PY{p}{:} \PY{l+s+s2}{\PYZdq{}}\PY{l+s+s2}{C}\PY{l+s+s2}{\PYZdq{}}
         \PY{p}{\PYZcb{}}\PY{p}{)}
         \PY{n}{g}\PY{o}{.}\PY{n}{influence\PYZus{}graph}\PY{p}{(}\PY{p}{)}
\end{Verbatim}
\end{small}

\begin{figure}[tbp]
    \centering
    \includegraphics[scale=0.7]{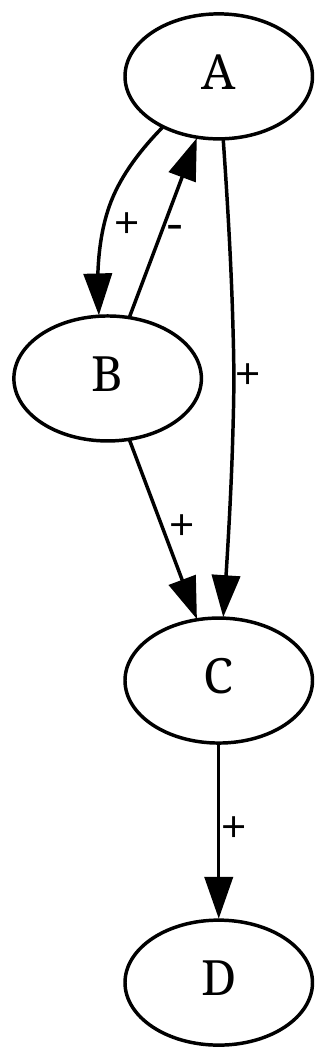}\caption{Influence graph of the example BN $g$ for P3}\label{fig:paper_files/paper_66_1.pdf}
    \end{figure}

    \noindent
    The resulting graphics is reproduced in Figure~\ref{fig:paper_files/paper_66_1.pdf}.

Unperturbed, this network has a single cyclic attractor, as \texttt{A}
and \texttt{B} are in a sustained oscillation.
\begin{small}
\begin{Verbatim}[commandchars=\\\{\}]
{\color{incolor}In [{\color{incolor}31}]:} \PY{n}{tabulate}\PY{p}{(}\PY{n}{g}\PY{o}{.}\PY{n}{attractors}\PY{p}{(}\PY{p}{)}\PY{p}{)}
\end{Verbatim}
\end{small}

\begin{small}
\begin{Verbatim}[commandchars=\\\{\}]
{\color{outcolor}Out[{\color{outcolor}31}]:}    A  B  C  D
         0  *  *  *  *
\end{Verbatim}
\end{small}
Enforcing that all attractors have \texttt{D} fixed to 1 can be achieved
either by perturbing \texttt{A}, or by perturbing only \texttt{C},
letting \texttt{A} and \texttt{B} oscillate.
\begin{small}
\begin{Verbatim}[commandchars=\\\{\}]
{\color{incolor}In [{\color{incolor}32}]:} \PY{n+nb}{list}\PY{p}{(}\PY{n}{marker\PYZus{}reprogramming}\PY{p}{(}\PY{n}{g}\PY{p}{,} \PY{p}{\PYZob{}}\PY{l+s+s2}{\PYZdq{}}\PY{l+s+s2}{D}\PY{l+s+s2}{\PYZdq{}}\PY{p}{:} \PY{l+m+mi}{1}\PY{p}{\PYZcb{}}\PY{p}{,} \PY{l+m+mi}{2}\PY{p}{)}\PY{p}{)}
\end{Verbatim}
\end{small}

\begin{small}
\begin{Verbatim}[commandchars=\\\{\}]
{\color{outcolor}Out[{\color{outcolor}32}]:} [\{'A': 1\}, \{'C': 1\}, \{'D': 1\}]
\end{Verbatim}
\end{small}
However, besides the forced activation of \texttt{A}, ensuring that all
and at least one fixed point have \texttt{D} active always requires
perturbing at least two components.
\begin{small}
\begin{Verbatim}[commandchars=\\\{\}]
{\color{incolor}In [{\color{incolor}33}]:} \PY{n+nb}{list}\PY{p}{(}\PY{n}{marker\PYZus{}reprogramming\PYZus{}fixpoints}\PY{p}{(}\PY{n}{g}\PY{p}{,} \PY{p}{\PYZob{}}\PY{l+s+s2}{\PYZdq{}}\PY{l+s+s2}{D}\PY{l+s+s2}{\PYZdq{}}\PY{p}{:} \PY{l+m+mi}{1}\PY{p}{\PYZcb{}}\PY{p}{,} \PY{l+m+mi}{2}\PY{p}{)}\PY{p}{)}
\end{Verbatim}
\end{small}

\begin{small}
\begin{Verbatim}[commandchars=\\\{\}]
{\color{outcolor}Out[{\color{outcolor}33}]:} [\{'A': 1\},
          \{'D': 1, 'A': 0\},
          \{'D': 1, 'B': 0\},
          \{'B': 1, 'D': 1\},
          \{'C': 1, 'A': 0\},
          \{'C': 1, 'B': 0\},
          \{'B': 1, 'C': 1\}]
\end{Verbatim}
\end{small}
\hypertarget{soure-marker-reprograming-of-attractors-p4}{%
\subsubsection{Soure-marker reprograming of attractors
(P4)}\label{soure-marker-reprograming-of-attractors-p4}}

Given an initial configuration \(z\), we identify the perturbations
\(P\) of at most \(k\) components so that the configurations of the all
the attractors of \(f/P\) that are reachable from \(z\) match with the
given marker \(M\) (i.e., in each reachable attractor, the specified
markers cannot oscillate). Thus, P4 is the same problem as P3, except
that we focus only on attractors reachable from \(z\), therefore
potentially requiring fewer perturbations.

The associated decision problem can be expressed as follows:

\begin{equation}
\exists P\in\mathbb M^k, \forall x\in\mathbb B^n, x\models M \vee x\notin\bar\rho^{f/P}_{\mathrm{mp}}(z)  \vee \neg\operatorname{IN-ATTRACTOR}_P(x)
\end{equation}

(``There exists a perturbation \(P\) of at most \(k\) components, such
that for all configurations \(x\), either \(x\) matches with the marker
\(M\), or \(x\) does not belong to an attractor, or \(x\) is not
reachable from \(z\)'').

By integrating the definition of the \(\operatorname{IN-ATTRACTOR}\)
property, we obtain the following \(\exists\forall\exists\)-expression:
\begin{equation}
\exists P\in\mathbb M^k, \forall x\in\mathbb B^n, x\models M \vee x\notin\bar\rho^{f/P}_{\mathrm{mp}}(z) \vee \exists y\in\mathbb B^n,
   y\in \operatorname{TS}_P(x), \operatorname{TS}_P(y) \neq \operatorname{TS}_P(x)
\end{equation}

The resolution of the problem can be approached in a very similar way to
P3, i.e., by solving its complement. The equation is almost the same,
with the addition that \(x\) must be reachable from \(z\), leading to
the \(\exists\forall\)-expression: \begin{equation}
\exists P\in\mathbb M^k, \exists x\in\mathbb B^n, x\in\operatorname{TS}_P(z) \wedge x\not\models M\wedge \forall y\in\mathbb B^n, y\in \operatorname{TS}_P(x) \implies \operatorname{TS}_P(y) \neq \operatorname{TS}_P(x)
\enspace.
\end{equation}

With the \emph{BoNesis} Python interface, this reprogramming property is
declared as follows, where \texttt{f} is a BN, \texttt{z} the initial
configuration, \texttt{M} the marker, and \texttt{k} the maximum number
of components that can be perturbed (at most \(n\)):
\begin{small}
\begin{Verbatim}[commandchars=\\\{\}]
{\color{incolor}In [{\color{incolor}34}]:} \PY{k}{def} \PY{n+nf}{source\PYZus{}marker\PYZus{}reprogramming}\PY{p}{(}\PY{n}{f}\PY{p}{:} \PY{n}{BooleanNetwork}\PY{p}{,}
                                         \PY{n}{z}\PY{p}{:} \PY{n+nb}{dict}\PY{p}{[}\PY{n+nb}{str}\PY{p}{,}\PY{n+nb}{bool}\PY{p}{]}\PY{p}{,}
                                         \PY{n}{M}\PY{p}{:} \PY{n+nb}{dict}\PY{p}{[}\PY{n+nb}{str}\PY{p}{,}\PY{n+nb}{bool}\PY{p}{]}\PY{p}{,}
                                         \PY{n}{k}\PY{p}{:} \PY{n+nb}{int}\PY{p}{)}\PY{p}{:}
             \PY{n}{bo} \PY{o}{=} \PY{n}{bonesis}\PY{o}{.}\PY{n}{BoNesis}\PY{p}{(}\PY{n}{f}\PY{p}{)}
             \PY{n}{coP} \PY{o}{=} \PY{n}{bo}\PY{o}{.}\PY{n}{Some}\PY{p}{(}\PY{n}{max\PYZus{}size}\PY{o}{=}\PY{n}{k}\PY{p}{)}
             \PY{k}{with} \PY{n}{bo}\PY{o}{.}\PY{n}{mutant}\PY{p}{(}\PY{n}{coP}\PY{p}{)}\PY{p}{:}
                 \PY{n}{x} \PY{o}{=} \PY{n}{bo}\PY{o}{.}\PY{n}{cfg}\PY{p}{(}\PY{p}{)}
                 \PY{n}{bo}\PY{o}{.}\PY{n}{in\PYZus{}attractor}\PY{p}{(}\PY{n}{x}\PY{p}{)}
                 \PY{n}{x} \PY{o}{!=} \PY{n}{bo}\PY{o}{.}\PY{n}{obs}\PY{p}{(}\PY{n}{M}\PY{p}{)}
                 \PY{o}{\PYZti{}}\PY{n}{bo}\PY{o}{.}\PY{n}{obs}\PY{p}{(}\PY{n}{z}\PY{p}{)} \PY{o}{\PYZgt{}}\PY{o}{=} \PY{n}{x}
             \PY{k}{return} \PY{n}{coP}\PY{o}{.}\PY{n}{complementary\PYZus{}assignments}\PY{p}{(}\PY{p}{)}
\end{Verbatim}
\end{small}

The above code snippet is very similar to the previous
\texttt{marker\_reprogramming}, with the addition of the
\texttt{\textasciitilde{}bo.obs(z)\ \textgreater{}=\ x} instruction
which declares that \texttt{x}, a configuration which belongs to an
attractor of the perturbed BN and which does not match with \texttt{M},
is reachable from \texttt{z}.
The corresponding command line is of the form

\begin{Shaded}
\begin{Highlighting}[]
\ExtensionTok{bonesis{-}reprogramming}\NormalTok{ model.bnet M k }\AttributeTok{{-}{-}reachable{-}from}\NormalTok{ z}
\end{Highlighting}
\end{Shaded}

where \texttt{model.bnet} is a BN encoded in BooleanNet format,
\texttt{M} specifies the marker as a JSON map, \texttt{k} is the maximum
number of perturbations, and \texttt{z} is the initial configuration as
a JSON map.
\paragraph{Example}

Let us consider again the BN \texttt{f} analyzed in the previous
section. By focusing only on attractors reachable from the configuration
where \texttt{A} is fixed to 1 and other nodes to 0, the reprogramming
required to make all attractors have \texttt{C} fixed to 1 consists only
of fixing \texttt{D} to 0. Note that in the specific example, the
reprogramming of reachable fixed point would give an equivalent result.
\begin{small}
\begin{Verbatim}[commandchars=\\\{\}]
{\color{incolor}In [{\color{incolor}35}]:} \PY{n}{z} \PY{o}{=} \PY{n}{f}\PY{o}{.}\PY{n}{zero}\PY{p}{(}\PY{p}{)}
         \PY{n}{z}\PY{p}{[}\PY{l+s+s2}{\PYZdq{}}\PY{l+s+s2}{A}\PY{l+s+s2}{\PYZdq{}}\PY{p}{]} \PY{o}{=} \PY{l+m+mi}{1}
         \PY{n+nb}{list}\PY{p}{(}\PY{n}{source\PYZus{}marker\PYZus{}reprogramming}\PY{p}{(}\PY{n}{f}\PY{p}{,} \PY{n}{z}\PY{p}{,} \PY{p}{\PYZob{}}\PY{l+s+s2}{\PYZdq{}}\PY{l+s+s2}{C}\PY{l+s+s2}{\PYZdq{}}\PY{p}{:} \PY{l+m+mi}{1}\PY{p}{\PYZcb{}}\PY{p}{,} \PY{l+m+mi}{3}\PY{p}{)}\PY{p}{)}
\end{Verbatim}
\end{small}

\begin{small}
\begin{Verbatim}[commandchars=\\\{\}]
{\color{outcolor}Out[{\color{outcolor}35}]:} [\{'D': 0\}, \{'C': 1\}]
\end{Verbatim}
\end{small}
The same results can be obtained using the command line as follows.
\begin{small}
\begin{Verbatim}[commandchars=\\\{\}]
{\color{incolor}In [{\color{incolor}36}]:} \PY{o}{!}bonesis\PYZhy{}reprogramming\PY{+w}{ }example3.bnet\PY{+w}{ }\PY{l+s+s1}{\PYZsq{}\PYZob{}\PYZdq{}C\PYZdq{}: 1\PYZcb{}\PYZsq{}}\PY{+w}{ }\PY{l+m}{3}\PY{+w}{ }\PY{err}{\PYZbs{}}
             \PY{o}{\PYZhy{}}\PY{o}{\PYZhy{}}\PY{n}{reachable}\PY{o}{\PYZhy{}}\PY{k+kn}{from} \PY{l+s+s1}{\PYZsq{}}\PY{l+s+s1}{\PYZob{}}\PY{l+s+s1}{\PYZdq{}}\PY{l+s+s1}{A}\PY{l+s+s1}{\PYZdq{}}\PY{l+s+s1}{: 1, }\PY{l+s+s1}{\PYZdq{}}\PY{l+s+s1}{B}\PY{l+s+s1}{\PYZdq{}}\PY{l+s+s1}{: 0, }\PY{l+s+s1}{\PYZdq{}}\PY{l+s+s1}{C}\PY{l+s+s1}{\PYZdq{}}\PY{l+s+s1}{: 0, }\PY{l+s+s1}{\PYZdq{}}\PY{l+s+s1}{D}\PY{l+s+s1}{\PYZdq{}}\PY{l+s+s1}{: 0\PYZcb{}}\PY{l+s+s1}{\PYZsq{}}
\end{Verbatim}
\end{small}

    \begin{Verbatim}[commandchars=\\\{\}]
\{'D': 0\}
\{'C': 1\}

    \end{Verbatim}

\hypertarget{reprogramming-of-ensembles-of-boolean-networks}{%
\subsection{Reprogramming of ensembles of Boolean
networks}\label{reprogramming-of-ensembles-of-boolean-networks}}

In the previous section, the reprogramming was performed on a single BN,
by giving to \emph{BoNesis} the singleton domain of BNs to consider. As
described in the Method section, \emph{BoNesis} can reason on ensembles
of BNs, either specified explicitly, or implicitly by an influence
graph. The functions defined above can then be directly applied to such
ensembles of BNs. In this section, we briefly discuss how the resulting
reprogramming solutions should then be interpreted with respect to these
ensembles.

Given a domain of BNs \(\mathbb F\), \emph{BoNesis} returns a solution
whenever at least one BN of this domain satisfies the given properties.
Intuitively, this means that the logic satisfiability problem is of the
form \(\exists f\in \mathbb F, \Phi(f)\). As detailed in
\citep{bn-synthesis-ICTAI19} in the scope of locally-monotone BNs, the
size of the ``\(f\in\mathbb F\)'' formula is, in general, exponential
(binomial coefficient) with the in-degree of nodes in the influence
graph. This complexity is due to the maximum number of clauses a Boolean
function can have. Our encoding in \emph{BoNesis} allows specifying an
upper bound to this number, which enables tackling very large scale
instances although giving access only to a subset of \(\mathbb F\). The
encoding of \(\mathbb F\) in \emph{BoNesis} also supports enforcing a
canonic representation of BNs in order to offer a non-redundant
enumeration of the BNs, at the price of a quadratic size of the formula.
However, in our case, as we are only interested in enumerating the
perturbations, the canonic form of BNs is not needed.

In the case of our implementation of marker reprogramming of fixed
points (P1 and P2), the expression becomes of the form:
\[\exists f\in\mathbb F, \exists P\in\mathbb M^{\leq k}, \cdots\]
Therefore, a perturbation \(P\) is returned as soon as it is a
reprogramming solution for \emph{at least one} BN of the input domain:
\(P\) may not work on every BN in \(\mathbb F\), but at least one.

In the case of our implementation of marker reprogramming of attractors
(P3 and P4), because we tackle the complementary problem, the expression
becomes of the form:
\[\exists f\in\mathbb F, \exists coP\in\mathbb M^{\leq k}, \cdots\]
Therefore, a \emph{bad} perturbation \(coP\) is returned as soon as it
is a bad perturbation for at least one BN of the input domain. By taking
the complement of these perturbations (in \(\mathbb M^{\leq k}\)), we
obtain that the returned perturbations are reprogramming solutions
\emph{for all} the BNs in \(\mathbb F\).
Let us illustrate the ensemble reprogramming with the following example.
First, let us define an influence graph to delimit the domain of
admissible BNs:
\begin{small}
\begin{Verbatim}[commandchars=\\\{\}]
{\color{incolor}In [{\color{incolor}37}]:} \PY{n}{dom} \PY{o}{=} \PY{n}{bonesis}\PY{o}{.}\PY{n}{InfluenceGraph}\PY{p}{(}\PY{p}{[}
             \PY{p}{(}\PY{l+s+s2}{\PYZdq{}}\PY{l+s+s2}{C}\PY{l+s+s2}{\PYZdq{}}\PY{p}{,} \PY{l+s+s2}{\PYZdq{}}\PY{l+s+s2}{B}\PY{l+s+s2}{\PYZdq{}}\PY{p}{,} \PY{p}{\PYZob{}}\PY{l+s+s2}{\PYZdq{}}\PY{l+s+s2}{sign}\PY{l+s+s2}{\PYZdq{}}\PY{p}{:} \PY{l+m+mi}{1}\PY{p}{\PYZcb{}}\PY{p}{)}\PY{p}{,}
             \PY{p}{(}\PY{l+s+s2}{\PYZdq{}}\PY{l+s+s2}{A}\PY{l+s+s2}{\PYZdq{}}\PY{p}{,} \PY{l+s+s2}{\PYZdq{}}\PY{l+s+s2}{C}\PY{l+s+s2}{\PYZdq{}}\PY{p}{,} \PY{p}{\PYZob{}}\PY{l+s+s2}{\PYZdq{}}\PY{l+s+s2}{sign}\PY{l+s+s2}{\PYZdq{}}\PY{p}{:} \PY{l+m+mi}{1}\PY{p}{\PYZcb{}}\PY{p}{)}\PY{p}{,}
             \PY{p}{(}\PY{l+s+s2}{\PYZdq{}}\PY{l+s+s2}{B}\PY{l+s+s2}{\PYZdq{}}\PY{p}{,} \PY{l+s+s2}{\PYZdq{}}\PY{l+s+s2}{C}\PY{l+s+s2}{\PYZdq{}}\PY{p}{,} \PY{p}{\PYZob{}}\PY{l+s+s2}{\PYZdq{}}\PY{l+s+s2}{sign}\PY{l+s+s2}{\PYZdq{}}\PY{p}{:} \PY{o}{\PYZhy{}}\PY{l+m+mi}{1}\PY{p}{\PYZcb{}}\PY{p}{)}\PY{p}{,}
             \PY{p}{(}\PY{l+s+s2}{\PYZdq{}}\PY{l+s+s2}{C}\PY{l+s+s2}{\PYZdq{}}\PY{p}{,} \PY{l+s+s2}{\PYZdq{}}\PY{l+s+s2}{D}\PY{l+s+s2}{\PYZdq{}}\PY{p}{,} \PY{p}{\PYZob{}}\PY{l+s+s2}{\PYZdq{}}\PY{l+s+s2}{sign}\PY{l+s+s2}{\PYZdq{}}\PY{p}{:} \PY{l+m+mi}{1}\PY{p}{\PYZcb{}}\PY{p}{)}\PY{p}{,}
         \PY{p}{]}\PY{p}{,} \PY{n}{exact}\PY{o}{=}\PY{k+kc}{True}\PY{p}{,} \PY{n}{canonic}\PY{o}{=}\PY{k+kc}{False}\PY{p}{)} \PY{c+c1}{\PYZsh{} we disable canonic encoding}
         \PY{n}{dom}
\end{Verbatim}
\end{small}

\begin{figure}[tbp]
    \centering
    \includegraphics[scale=0.7]{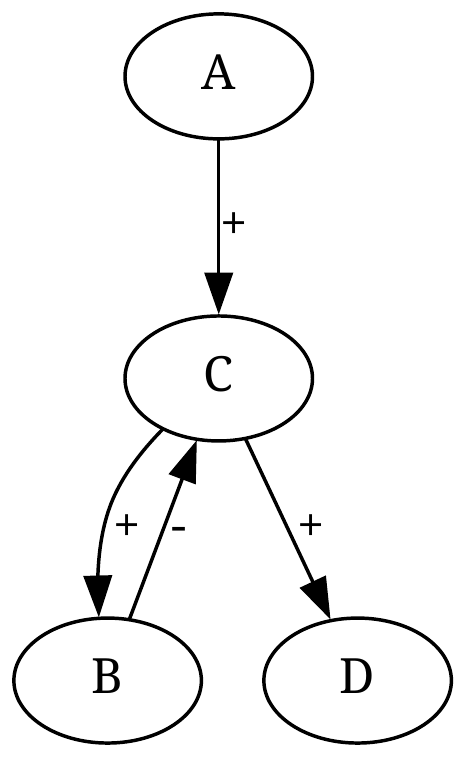}\caption{Influence graph delimiting the ensembles of BNs in the example}\label{fig:paper_files/paper_83_1.pdf}
    \end{figure}

    \noindent
    The resulting graphics is reproduced in Figure~\ref{fig:paper_files/paper_83_1.pdf}.

This domain encloses all the BNs having exactly (\texttt{exact=True})
the specified influence graph, 4 distinct BNs in this case:
\begin{small}
\begin{Verbatim}[commandchars=\\\{\}]
{\color{incolor}In [{\color{incolor}38}]:} \PY{n}{dom}\PY{o}{.}\PY{n}{canonic} \PY{o}{=} \PY{k+kc}{True} \PY{c+c1}{\PYZsh{} we set canonic encoding for enumerating BNs}
         \PY{n}{F} \PY{o}{=} \PY{n+nb}{list}\PY{p}{(}\PY{n}{bonesis}\PY{o}{.}\PY{n}{BoNesis}\PY{p}{(}\PY{n}{dom}\PY{p}{)}\PY{o}{.}\PY{n}{boolean\PYZus{}networks}\PY{p}{(}\PY{p}{)}\PY{p}{)}
         \PY{n}{dom}\PY{o}{.}\PY{n}{canonic} \PY{o}{=} \PY{k+kc}{False}
         \PY{n}{pd}\PY{o}{.}\PY{n}{DataFrame}\PY{p}{(}\PY{n}{F}\PY{p}{)}
\end{Verbatim}
\end{small}

\begin{small}
\begin{Verbatim}[commandchars=\\\{\}]
{\color{outcolor}Out[{\color{outcolor}38}]:}    A  B     C  D
         0  1  C  A|!B  C
         1  0  C  A|!B  C
         2  0  C  A\&!B  C
         3  1  C  A\&!B  C
\end{Verbatim}
\end{small}
Let us explore the attractors of each individual BNs:
\begin{small}
\begin{Verbatim}[commandchars=\\\{\}]
{\color{incolor}In [{\color{incolor}39}]:} \PY{k}{for} \PY{n}{i}\PY{p}{,} \PY{n}{f} \PY{o+ow}{in} \PY{n+nb}{enumerate}\PY{p}{(}\PY{n}{F}\PY{p}{)}\PY{p}{:}
             \PY{n+nb}{print}\PY{p}{(}\PY{l+s+sa}{f}\PY{l+s+s2}{\PYZdq{}}\PY{l+s+s2}{Attractors of BN }\PY{l+s+si}{\PYZob{}}\PY{n}{i}\PY{l+s+si}{\PYZcb{}}\PY{l+s+s2}{:}\PY{l+s+s2}{\PYZdq{}}\PY{p}{,} \PY{n+nb}{list}\PY{p}{(}\PY{n}{f}\PY{o}{.}\PY{n}{attractors}\PY{p}{(}\PY{p}{)}\PY{p}{)}\PY{p}{)}
\end{Verbatim}
\end{small}

    \begin{Verbatim}[commandchars=\\\{\}]
Attractors of BN 0: [\{'A': 1, 'B': 1, 'C': 1, 'D': 1\}]
Attractors of BN 1: [\{'A': 0, 'B': '*', 'C': '*', 'D': '*'\}]
Attractors of BN 2: [\{'A': 0, 'B': 0, 'C': 0, 'D': 0\}]
Attractors of BN 3: [\{'A': 1, 'B': '*', 'C': '*', 'D': '*'\}]

    \end{Verbatim}

In this example, we focus on reprogramming the attractors so that the
component \texttt{D} is fixed to 1.

On the one hand, when reprogramming fixed points only, because one BN
already verifies this property, the empty perturbation is a solution:
\begin{small}
\begin{Verbatim}[commandchars=\\\{\}]
{\color{incolor}In [{\color{incolor}40}]:} \PY{n+nb}{list}\PY{p}{(}\PY{n}{marker\PYZus{}reprogramming\PYZus{}fixpoints}\PY{p}{(}\PY{n}{dom}\PY{p}{,} \PY{p}{\PYZob{}}\PY{l+s+s2}{\PYZdq{}}\PY{l+s+s2}{D}\PY{l+s+s2}{\PYZdq{}}\PY{p}{:} \PY{l+m+mi}{1}\PY{p}{\PYZcb{}}\PY{p}{,} \PY{l+m+mi}{2}\PY{p}{)}\PY{p}{)}
\end{Verbatim}
\end{small}

\begin{small}
\begin{Verbatim}[commandchars=\\\{\}]
{\color{outcolor}Out[{\color{outcolor}40}]:} [\{\}]
\end{Verbatim}
\end{small}
On the other hand, the reprogramming of attractors returns solution that
work on every BN:
\begin{small}
\begin{Verbatim}[commandchars=\\\{\}]
{\color{incolor}In [{\color{incolor}41}]:} \PY{n+nb}{list}\PY{p}{(}\PY{n}{marker\PYZus{}reprogramming}\PY{p}{(}\PY{n}{dom}\PY{p}{,} \PY{p}{\PYZob{}}\PY{l+s+s2}{\PYZdq{}}\PY{l+s+s2}{D}\PY{l+s+s2}{\PYZdq{}}\PY{p}{:} \PY{l+m+mi}{1}\PY{p}{\PYZcb{}}\PY{p}{,} \PY{l+m+mi}{2}\PY{p}{)}\PY{p}{)}
\end{Verbatim}
\end{small}

\begin{small}
\begin{Verbatim}[commandchars=\\\{\}]
{\color{outcolor}Out[{\color{outcolor}41}]:} [\{'C': 1\}, \{'D': 1\}, \{'B': 0, 'A': 1\}]
\end{Verbatim}
\end{small}
Indeed, fixed \texttt{C} to 1, ensures in each case that \texttt{D} is
fixed to 1.
The computation of universal solutions for the reprogramming of fixed
points can be tackled by following a similar encoding than the
reprogramming of attractors, i.e., by identifying perturbations which do
not fulfill the property for at least one BN in the domain (the
complement results in perturbations working for all the BNs):
\begin{small}
\begin{Verbatim}[commandchars=\\\{\}]
{\color{incolor}In [{\color{incolor}42}]:} \PY{k}{def} \PY{n+nf}{universal\PYZus{}marker\PYZus{}reprogramming\PYZus{}fixpoints}\PY{p}{(}\PY{n}{f}\PY{p}{:} \PY{n}{BooleanNetwork}\PY{p}{,}
                                                      \PY{n}{M}\PY{p}{:} \PY{n+nb}{dict}\PY{p}{[}\PY{n+nb}{str}\PY{p}{,}\PY{n+nb}{bool}\PY{p}{]}\PY{p}{,}
                                                      \PY{n}{k}\PY{p}{:} \PY{n+nb}{int}\PY{p}{)}\PY{p}{:}
             \PY{n}{bo} \PY{o}{=} \PY{n}{bonesis}\PY{o}{.}\PY{n}{BoNesis}\PY{p}{(}\PY{n}{f}\PY{p}{)}
             \PY{n}{coP} \PY{o}{=} \PY{n}{bo}\PY{o}{.}\PY{n}{Some}\PY{p}{(}\PY{n}{max\PYZus{}size}\PY{o}{=}\PY{n}{k}\PY{p}{)}
             \PY{k}{with} \PY{n}{bo}\PY{o}{.}\PY{n}{mutant}\PY{p}{(}\PY{n}{coP}\PY{p}{)}\PY{p}{:}
                 \PY{n}{x} \PY{o}{=} \PY{n}{bo}\PY{o}{.}\PY{n}{cfg}\PY{p}{(}\PY{p}{)}
                 \PY{n}{bo}\PY{o}{.}\PY{n}{fixed}\PY{p}{(}\PY{n}{x}\PY{p}{)} \PY{c+c1}{\PYZsh{} x is a fixed point}
                 \PY{n}{x} \PY{o}{!=} \PY{n}{bo}\PY{o}{.}\PY{n}{obs}\PY{p}{(}\PY{n}{M}\PY{p}{)} \PY{c+c1}{\PYZsh{} x does not match with M}
             \PY{k}{return} \PY{n}{coP}\PY{o}{.}\PY{n}{complementary\PYZus{}assignments}\PY{p}{(}\PY{p}{)}
\end{Verbatim}
\end{small}

\begin{small}
\begin{Verbatim}[commandchars=\\\{\}]
{\color{incolor}In [{\color{incolor}43}]:} \PY{n+nb}{list}\PY{p}{(}\PY{n}{universal\PYZus{}marker\PYZus{}reprogramming\PYZus{}fixpoints}\PY{p}{(}\PY{n}{dom}\PY{p}{,} \PY{p}{\PYZob{}}\PY{l+s+s2}{\PYZdq{}}\PY{l+s+s2}{D}\PY{l+s+s2}{\PYZdq{}}\PY{p}{:} \PY{l+m+mi}{1}\PY{p}{\PYZcb{}}\PY{p}{,} \PY{l+m+mi}{2}\PY{p}{)}\PY{p}{)}
\end{Verbatim}
\end{small}

\begin{small}
\begin{Verbatim}[commandchars=\\\{\}]
{\color{outcolor}Out[{\color{outcolor}43}]:} [\{'C': 1\}, \{'A': 1\}, \{'D': 1\}]
\end{Verbatim}
\end{small}
Note that in this implementation, we do not ensure the existence of a
fixed point after reprogramming. This is why the perturbation fixing
only \texttt{A} to 1 is considered as a solution in our example:
\begin{small}
\begin{Verbatim}[commandchars=\\\{\}]
{\color{incolor}In [{\color{incolor}44}]:} \PY{k}{for} \PY{n}{i}\PY{p}{,} \PY{n}{f} \PY{o+ow}{in} \PY{n+nb}{enumerate}\PY{p}{(}\PY{n}{F}\PY{p}{)}\PY{p}{:}
             \PY{n}{f}\PY{p}{[}\PY{l+s+s2}{\PYZdq{}}\PY{l+s+s2}{A}\PY{l+s+s2}{\PYZdq{}}\PY{p}{]} \PY{o}{=} \PY{l+m+mi}{1}
             \PY{n+nb}{print}\PY{p}{(}\PY{l+s+sa}{f}\PY{l+s+s2}{\PYZdq{}}\PY{l+s+s2}{Attractors of BN }\PY{l+s+si}{\PYZob{}}\PY{n}{i}\PY{l+s+si}{\PYZcb{}}\PY{l+s+s2}{ after fixing A to 1:}\PY{l+s+s2}{\PYZdq{}}\PY{p}{,} \PY{n+nb}{list}\PY{p}{(}\PY{n}{f}\PY{o}{.}\PY{n}{attractors}\PY{p}{(}\PY{p}{)}\PY{p}{)}\PY{p}{)}
\end{Verbatim}
\end{small}

    \begin{Verbatim}[commandchars=\\\{\}]
Attractors of BN 0 after fixing A to 1: [\{'A': 1, 'B': 1, 'C': 1, 'D': 1\}]
Attractors of BN 1 after fixing A to 1: [\{'A': 1, 'B': 1, 'C': 1, 'D': 1\}]
Attractors of BN 2 after fixing A to 1: [\{'A': 1, 'B': '*', 'C': '*', 'D': '*'\}]
Attractors of BN 3 after fixing A to 1: [\{'A': 1, 'B': '*', 'C': '*', 'D': '*'\}]
    \end{Verbatim}
\noindent
As BNs 2 and 3 have no fixed point, they fulfill the criteria ``all the
fixed points match with marker \texttt{M}''.
\hypertarget{scalability}{%
\subsection{Scalability}\label{scalability}}

In order to evaluate the scalability on realistic BNs, we use the
benchmark constituted by \citet{Moon22} to evaluate the reprogramming of
fixed points (P1). Their benchmark gathers 10 locally-monotone BNs and 1
non-monotone one, that BoNesis cannot address. The dimension of the 10
BNs are respectively 14, 17, 18, 20, 28, 32, 53, 59, 66, and 75. For
each of these models, a target marker for reprogramming has been defined
from the corresponding published studies. Moreover, a subset of nodes
has been declared as uncontrollable to avoid trivial solutions. We used
this benchmark to evaluate the scalability of the P1 and P3
implementation we propose in this paper.

For these 10 models, we applied the P1 and P3 reprogramming for
different maximum number of simultaneous perturbations (denoted \(k\) in
the previous sections), up to \(k=6\). In each case, we measured the
time for the first solution, for listing up to 100 solutions, and for
listing all the solutions, with a timeout of 10 minutes. The experiments
have been performed on an Intel(R) Xeon(R) processor at 3.3Ghz with 16BG
of RAM.\\
In the case of P1, with \(k=6\), it took around 1s to get at least one
solution for each of the 10 models; up to 100 solutions have been listed
in the same timing, except for one model which took 8s. The full listing
of solutions of 3 of the larger models have timed out, the rest
necessitating between 1 and 18s. With \(k=4\), \emph{BoNesis} was able
to list all the solutions for all models (up to 5min for one of the
larger model).\\
In the case of P3, with \(k\geq 4\), 3 of the 10 models could not find a
single solution in the given time limit; for most of the other models, a
first solution was found in around 1s, a couple of models took around
1-2min. The enumeration of the first 100 solutions took a similar time
with \(k=4\), but timed out with \(k=6\) for all but the 4 smallest
models. With \(k\leq 2\), \emph{BoNesis} has found all the solutions to
P3 for all the 10 models in a few seconds maximum.

These experiments testify of the difference of complexity between P1 and
P3, and more precisely on the resolution approach taken for P3: the
computation of the complementary sets of solutions becomes rapidly
intractable for large combinations of perturbation. Indeed, in practice,
there are many bad perturbations, thus their enumeration, which is
necessary to compute their complement, is an important bottleneck.

Evaluating the scalability of P2 and P4 would require defining initial
configurations which are meaningful for the different models, which are
not available in the selected benchmark. Nevertheless, because the
source constraint does not change the complexity classes, we can
conjecture that their scalability should be comparable to P1 and P3
respectively. Moreover, having benchmarks at larger scale would be
insightful, but none of them are available to the best of our knowledge.
It should be noted \emph{BoNesis} has been applied to do BN synthesis
for models with 1,000 nodes \citep{bn-synthesis-ICTAI19}, suggesting a
potential applicability of \emph{BoNesis} for the reprogramming of large
BNs.

As stressed in the introduction, there exists only tools addressing P1
to compare with. The experiments of \citet{Moon22} show that their
bilevel integer programming-based method systematically outperforms the
ASP implementation of pyActoNet \citep{Biane2018}. On the same
benchmark, \emph{BoNesis} performed either similarly or in shorter time,
albeit limited to locally-monotone BNs only.
\hypertarget{discussion}{%
\section{Discussion}\label{discussion}}

In this paper, we demonstrated how the \emph{BoNesis} Python library can
be employed to fully characterize permanent perturbations which
guarantee that all the fixed points or all minimal trap spaces of the
perturbed BN have a subset of components fixed to desired values
(marker). We focused on reprogramming for achieving elementary dynamical
properties, that are the fixed points or attractors, optionally
reachable from a given configuration. Nevertheless, the snippets shown
in this paper can be extended to account for more complex or specific
dynamical properties after mutation, e.g., existence of additional
trajectories, considering multiple initial configurations.

It should be noted that the candidate combinations of perturbations are
computed solely based on the Boolean dynamics, and do not account for
experimental feasibility, e.g., in the scope of models of biological
systems. Future work may consider optimization or prioritization of
perturbations based on such extra information. Currently, \emph{BoNesis}
enables specifying uncontrollable components which must not be perturbed
(\texttt{exclude} option for the \texttt{Some} object, or
\texttt{-\/-exclude} for the command line, taking a list of components
which should be excluded from the candidate perturbations).

We considered for problems, referred to as P1, P2, P3, P4, where P1-P2
relate to the reprogramming of fixed points, and P3-P4 to the
reprogramming of MP attractors (i.e., minimal trap spaces). The
computational complexity of P1-P2 allows an efficient implementation
using Answer-Set Programming (ASP), whereas the one of P3-P4 necessitate
working around complementary solution to fit into the expressiveness of
ASP, limiting their scalability. Future work may explore alternative
implementations using different logic frameworks.

The identified perturbations may destroy and create new fixed points and
attractors for the BN. This is a significant difference with most of the
methods developed in the literature where many focus on the control
towards attractors of the unperturbed BNs only. Whereas the problem P1
which has been already addressed with different methods, we are not
aware of any other approach tackling P2, P3, and P4.

Besides the four reprogramming problems tackled in this paper, an
additional variant would be the marker-reprogramming of fixed points
which also ensures the absence of cyclic attractors. Note that its
complexity is equivalent to the one of P3/P4, i.e., it can be expressed
as a \(\exists\forall\exists\)-expression. This problem may be relevant
for modeling cases where cyclic attractors do not make sense. The
programming interface of \emph{BoNesis} do not permit an efficient
encoding of this problem at the moment.

This paper focused on permanent perturbations, i.e., enforcing the value
of one or several components constantly over time, independently of the
state of the system. \emph{Sequential} reprogramming
\citep{Mandon2017,Pardo2021} consists in applying sets of perturbations
at different time. This can lead to reducing the overall number of
component to perturb, by taking advantage of the natural transient
dynamics of the system. Sequential reprogramming brings the BN
reprogramming settings closer to classical control theory, as the
control can depend both on time and state of the system. Having fixed a
number of steps, say \(m\), the reprogramming problems consists in
identifying \(m\) sets of perturbations which will be applied in
sequence, and their application may be restricted to attractors only
\citep{Mandon2019}. Interestingly in that case, having fixed the number
of reprogramming steps, the computational complexity remains identical
to the one-step reprogramming with locally-monotone BNs, due to the
PTIME complexity of the reachability. For instance, the 2-steps
reprogramming of BNs along fixed points only with the MP update mode can
be expressed as the following \(\exists\forall\)-expression, as P1:

\[\exists P,Q\in\mathbb M^{\leq k}, \forall x,y\in\mathbb B^n,
   f_P(x) = x \Rightarrow
      \operatorname{reach}_Q(x,y) \Rightarrow
      f_Q(y) = y \Rightarrow y\models M\]

``There exist two sets of perturbations \(P\) and \(Q\), such that for
any configuration \(x\) and any configuration \(y\), if \(x\) is a fixed
point of under the perturbation \(P\), then if \(y\) is reachable from
\(x\) under the perturbation \(Q\), then if \(y\) is a fixed point under
the perturbation \(Q\), then it must match with the marker \(M\)''. The
more general 2-steps reprogramming along attractors can be expressed as
follows:

\[\exists P,Q\in\mathbb M^{\leq k}, \forall x,y\in\mathbb B^n,
\operatorname{IN-ATTRACTOR}_P(x) \Rightarrow
     \operatorname{reach}_Q(x,y) \Rightarrow
     \operatorname{IN-ATTRACTOR}_Q(y) \Rightarrow y \models M\]
Accounting for \(\operatorname{IN-ATTRACTOR}\), this leads to an
\(\exists\forall\exists\)-expression, as for the single-step
reprogramming. Future work may then investigate the encoding of
sequential reprogramming with \emph{BoNesis}.

Finally, we demonstrated how \emph{BoNesis} can be employed to reason on
the reprogramming of BNs, leading to either solutions that work for at
least one BN of the ensemble, or working on each of them (universal
reprogramming). We believe that reasoning on ensemble of models is a
promising direction to address the robustness of predictions in the
scope of applications in systems biology.
\hypertarget{acknowledgements}{%
\subsubsection{Acknowledgements}\label{acknowledgements}}

The author would like to thank Kyungduk Moon and Kangbok Lee for
stimulating discussions and for providing the model and configuration
files for the benchmarks of \citet{Moon22}.

\hypertarget{software-and-data-availability}{%
\subsubsection{Software and data
availability}\label{software-and-data-availability}}

The software \emph{BoNesis} is available at
\href{https://github.com/bnediction/bonesis}{github.com/bnediction/bonesis}.
The code of this paper uses version 0.4.93 (archived at
\href{https://doi.org/10.5281/zenodo.7657487}{doi:10.5281/zenodo.7657487}).
The notebook, models, and instructions for reproducing the results are
available at
\href{https://github.com/bnediction/reprogramming-with-bonesis}{github.com/bnediction/reprogramming-with-bonesis}
(archived at
\href{https://doi.org/10.5281/zenodo.7733095}{doi:10.5281/zenodo.7733095}).

\hypertarget{funding}{%
\subsubsection{Funding}\label{funding}}

This work has been supported by the French Agence Nationale pour la
Recherche (ANR) in the scope of the project
\href{https://bnediction.github.io}{BNeDiction} (grant number
ANR-20-CE45-0001).

\bibliographystyle{abbrvnat}
\bibliography{references}

\end{document}